\documentclass[letterpaper,twocolumn,10pt]{article}

\usepackage{usenix2019_v3}

\usepackage{url}
\usepackage{graphicx}
\usepackage{wrapfig}
\usepackage{sidecap}
\usepackage{color}
\usepackage{xcolor}
\usepackage{pifont}
\usepackage{upquote}
\usepackage{listings}
\usepackage{hyperref}
\usepackage{algorithm2e}
\usepackage{subcaption}
\usepackage{picins}
\usepackage[export]{adjustbox}
\hypersetup{breaklinks=true,
            bookmarks=true,
            pdfauthor={},
            colorlinks=true,
            citecolor=blue,
            urlcolor=blue,
            linkcolor=magenta,
            pdfborder={0 0 0}}
\definecolor{editorGray}{rgb}{0.95, 0.95, 0.95}
\definecolor{editorOcher}{rgb}{1, 0.5, 0} %
\definecolor{editorGreen}{rgb}{0, 0.5, 0} %
\definecolor{btq}{rgb}{0.03, 0.91, 0.87} %
\definecolor{dtq}{rgb}{0.0, 0.81, 0.82} %
\definecolor{cdb}{rgb}{0.37, 0.62, 0.63} %
\lstdefinelanguage{js}{
  morekeywords={typeof, new, true, false, catch, function, return, null, catch, switch, var, if, in, while, do, else, case, break},
  keywordstyle=\color{purple}\ttfamily,
  keywordstyle=[3]\color{orange},
  keywords=[3]{SBX, PROC},
  ndkeywords={class, export, require, boolean, throw, implements, this},
  ndkeywordstyle=\color{cyan}\ttfamily, %
  identifierstyle=\color{black},
  sensitive=false,
  comment=[l]{//},
  commentstyle=\color{lightgray},
  morecomment=[s]{/**, */},
  stringstyle=\color{blue}\ttfamily,
  morestring=[b]',
  morestring=[b]",
}

\lstdefinelanguage{es}{
  morekeywords={string, number, bool, p, v, e, x},
  keywordstyle=\color{darkgray},
  keywordstyle=[3]\color{orange},
  keywords=[3]{SBX, PROC},
  ndkeywords={undefined, null, delete},
  ndkeywordstyle=\color{black}\ttfamily\bfseries,
  identifierstyle=\color{black},
  sensitive=false,
  comment=[l]{//},
  commentstyle=\color{lightgray},
  stringstyle=\color{darkgray}\ttfamily,
  morestring=[b]',
  morestring=[b]",
  abovecaptionskip=0em,
  aboveskip=0em,
  belowcaptionskip=0em,
  belowskip=0em,
  frame=none,                     %
}

\lstdefinelanguage{Haskell}{
  morekeywords={match, with, end, where, map},
  keywordstyle=\color{purple}\ttfamily,
  ndkeywords={Object, Array, Fun, String, Bool, Undef, Null, Number},
  ndkeywordstyle=\color{cyan}\ttfamily, %
  identifierstyle=\color{black},
  sensitive=false,
  comment=[l]{//},
  commentstyle=\color{lightgray},
  morecomment=[s]{/**, */},
  stringstyle=\color{blue}\ttfamily,
  morestring=[b]',
  morestring=[b]",
  backgroundcolor=\color{white},   %
  basicstyle=\small\ttfamily,  %
  upquote=true,
  captionpos=b,                    %
  frame=single,                    %
  numbers=left,                    %
  numbersep=5pt,                   %
  xleftmargin=2em,
  numberstyle=\tiny\color{gray},   %
  rulecolor=\color{black},         %
}

\usepackage{soul}
\usepackage{xspace}
\usepackage{booktabs}
\usepackage{amsmath}

\usepackage{caption}
\SetAlCapNameFnt{\footnotesize}
\SetAlCapFnt{\footnotesize}

\usepackage{enumitem}
\setlist{noitemsep,leftmargin=10pt,topsep=2pt,parsep=2pt,partopsep=2pt}

\usepackage{cleveref}

\def\myomit#1{}
\def\eg{{\em e.g.}, }
\def\ie{{\em i.e.}, }
\def\etc{{\em etc.}\xspace}

\newcommand{\heading}[1]{\vspace{2pt}\noindent\textbf{#1}\enspace}
\newcommand{\ttt}[1]{\texttt{#1}}

\newcommand{\tcn}[1]{}

\newcommand{\new}{{\scshape{}Mir}\xspace} %
\newcommand{\sys}{\new}
\newcommand{\R}{\ttt{R}\xspace}
\newcommand{\W}{\ttt{W}\xspace}
\newcommand{\X}{\ttt{X}\xspace}

\newcommand{\I}{\ttt{I}\xspace}
\newcommand{\rwx}{\ttt{RWX}\xspace}

\newcommand{\sx}[1]{(\S\ref{#1})}
\newcommand{\cf}[1]{(\emph{Cf}.\S\ref{#1})}

\newcommand{\str}{\textbf{\ttt{*}}\xspace}
\newcommand{\ctx}[1]{\textbf{\textsf{\small \color{violet}#1}\xspace}}
\newcommand{\ctxt}[1]{\textbf{\textsf{\scriptsize \color{violet}#1}\xspace}}
\newcommand{\fld}[1]{\textbf{\texttt{\small \color{olive}#1}\xspace}}
\newcommand{\fldt}[1]{\textbf{\texttt{\scriptsize \color{olive}#1}\xspace}}

\newcommand{\topThousandpkgsToConsider}{986}
\newcommand{\topThousandAnalysisTime}{42 minutes}
\newcommand{\topThousandAnalysisTimePerLib}{2.5 seconds}

\newcommand{\topThousandSLOC}{5,826,357}
\newcommand{\topThousandPermPerLOC}{10}
\newcommand{\topThousandReadsMin}{0}
\newcommand{\topThousandReadsMax}{27736}

\newcommand{\topThousandReadsFirstQ}{13}
\newcommand{\topThousandReadsMedian}{39}
\newcommand{\topThousandReadsThirdQ}{156}
\newcommand{\topThousandWritesMin}{0}
\newcommand{\topThousandWritesMax}{9018}

\newcommand{\topThousandWritesFirstQ}{1}
\newcommand{\topThousandWritesMedian}{4}
\newcommand{\topThousandWritesThirdQ}{19}
\newcommand{\topThousandExecutesMin}{0}
\newcommand{\topThousandExecutesMax}{16638}

\newcommand{\topThousandExecutesFirstQ}{6}
\newcommand{\topThousandExecutesMedian}{20}
\newcommand{\topThousandExecutesThirdQ}{78}
\newcommand{\topThousandImportsMin}{0}
\newcommand{\topThousandImportsMax}{16639}

\newcommand{\topThousandImportsFirstQ}{2}
\newcommand{\topThousandImportsMedian}{6}
\newcommand{\topThousandImportsThirdQ}{29}
\newcommand{\topThousandTotalPermsMin}{0}
\newcommand{\topThousandTotalPermsMax}{66559}

\newcommand{\topThousandTotalPermsFirstQ}{25}
\newcommand{\topThousandTotalPermsMedian}{73}
\newcommand{\topThousandTotalPermsThirdQ}{295}
\newcommand{\topThousandPercReads}{50.33\%}
\newcommand{\topThousandPercExecutes}{22.97\%}

\begin{document}
\title{\sys: Automated Quantifiable Privilege Reduction\\ Against Dynamic Library Compromise in JavaScript}
\author{
  {\rm \normalsize Nikos Vasilakis\qquad Cristian-Alexandru Staicu$^{\dagger}$\qquad Greg Ntousakis$^{\circ}$\qquad Konstantinos Kallas$^{\ddagger}$\qquad }\\
  {\rm \normalsize Ben Karel$^{\diamond}$ \qquad Andr\'{e} DeHon$^{\ddagger}$ \qquad Michael Pradel$^{\dagger\dagger}$ }\\
  \small
   MIT, CSAIL \quad $^{\dagger}$TU Darmstadt \& CISPA \quad $^{\circ}$TU Crete \quad $^{\ddagger}$University of Pennsylvania \quad
   $^{\diamond}$Aarno Labs \quad $^{\dagger\dagger}$University of Stuttgart
}

\maketitle

\begin{abstract}
Third-party libraries ease the development of large-scale software systems. 
However, they often execute with significantly more privilege than needed to complete their task.
This additional privilege is often exploited at runtime via dynamic compromise, even when these libraries are not actively malicious.
\new addresses this problem by introducing a fine-grained read-write-execute (\rwx) permission model at the boundaries of libraries.
Every field of an imported library is governed by a set of permissions, which developers can express when importing libraries.
To enforce these permissions during program execution, \new transforms libraries and their context to add runtime checks.
As permissions can overwhelm developers, \sys's permission inference generates default permissions by analyzing how libraries are used by their consumers.
Applied to 50 popular libraries, \sys's prototype for JavaScript demonstrates that the \rwx permission model combines simplicity with power:
  it is simple enough to automatically infer 99.33\% of required permissions,
  it is expressive enough to defend against 16 real threats,
  it is efficient enough to be usable in practice (1.93\% overhead),
  and it enables a novel quantification of privilege reduction.
\end{abstract}

\section{Introduction}\label{introduction}
\label{intro}

Modern software development relies heavily on third-party libraries.\footnote{
  This paper uses the terms library, module, and package interchangeably.
  It also uses \ttt{import} everywhere, for both Node's \ttt{require} and ES6's \ttt{import}.
}
Such reliance has led to an explosion of attacks~\cite{maass2016theory, lauinger2017thou, long2015owasp, cadariu2015tracking, snyk, npmstudy:19}:
  overprivileged code in imported libraries provides an attack vector that is exploitable long after libraries reach their end-users.
Even when libraries are created and authored with the best possible intentions---\ie are not actively malicious---their privilege can be exploited at runtime to compromise the entire application---or worse, the broader system on which the application is executing.

Such \emph{dynamic compromise}---as opposed to compromise achieved earlier in the supply chain, for actively-malicious libraries---is possible due to several compounding factors.
Libraries offer a great deal of functionality, but only a small fraction of this functionality may be used by any one particular client~\cite{dynJS:15}.
Default-allow semantics give any library unrestricted access to all of the features in a programming language---\eg accessing global variables, introspecting and rewriting core functionality, and even importing other libraries~\cite{bittau2008wedge}.
A library's intended use is a small space well-understood by its developers, but unexpected or pathological use covers a much larger space, typically understood by the library's clients~\cite{buxton1970software}.
For example, only the client of a de-serialization library knows whether it will be fed non-sanitized input coming directly from the network;
  the library's developers cannot make such assumptions about its use.

\begin{figure}[t]
\includegraphics[width=0.49\textwidth, left]{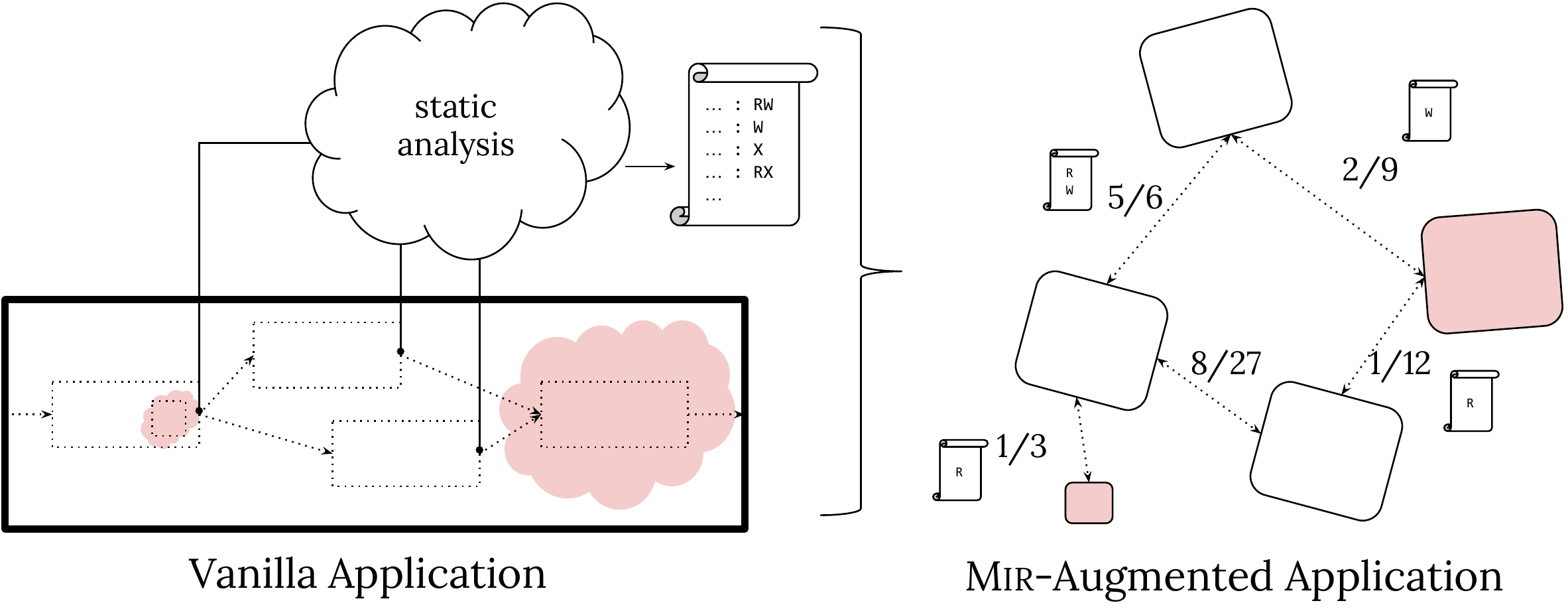}
\caption{
  \textbf{\sys overview.}
  \sys analyzes individual libraries, many of which may be subvertible, to generate permissions.
  It then enforces these permissions at runtime in order to (quantifiably) lower the privielge of these libraries over the rest of the application and its surrounding environment.
	\vspace{-5mm}
}
\label{fig:overview}
\end{figure}

To address dynamic compromise, \sys augments a module system with a model for specifying, enforcing, inferring, and quantifying the privilege available to libraries in a backward-compatible fashion (Fig.~\ref{fig:overview}).
\sys's key insight is that libraries cannot be subverted at runtime to exploit functionality to which they do not already have access.
Coupling default-deny semantics with explicit and statically-inferrable whitelisting, \sys minimizes the effects of dynamic compromise---regardless of the behavior exercised by a library and its clients.

\heading{Specification}
\sys's permission model allows specifying fine-grained \emph{read-write-execute permissions} (\rwx) that guard access to individual object fields available within the scope of a library.
Such objects and fields are available through both explicit library imports and built-in language features---\eg globals, process arguments, \ctx{import} capabilities.
In the aforementioned de-serialization library, two example permissions would be 
  (1) \R only on \ttt{process.env.PWD}, only allowing read-access to the \ttt{PWD} environment variable,
  and
  (2) \X only on \ttt{serialize}, disallowing execute-access to the---unused but exploitable---de-serialization function.

\heading{Inference}
Unfortunately, manually specifying permissions is challenging, even for security-paranoid developers.
Typical challenges include
  (1) naming issues, such as variable aliasing, poor variable names, and non-local use, 
  (2) library-internal code, possibly not intended for humans, and 
  (3) continuous codebase evolution, which requires updating the specification for every code change.
To address these problems, \sys analyzes name uses within a library to automatically infer permissions. %
The majority of the analysis is static, augmented with a short phase of import-time analysis addressing common runtime meta-programming concerns.

\heading{Quantification}
\sys improves security by restricting the permissions granted to libraries by default; 
  at the same time, inferrring 100\% of a library's permissions is intractable for dynamic languages such as JavaScript.
To address this conundrum, we propose a quantitative privilege reduction metric to
evaluate the permission prevention that \sys exercises.
This quantification is achieved by comparing the permissions granted by \sys to those a library would have by default---\ie by statically counting all the names available in a library's lexical scope.

\heading{Enforcement}
To enforce \rwx permissions, \new transforms libraries to add security monitors.
\sys's transformations are enabled by the fact that JavaScript features a library-import mechanism that loads code at runtime as a string.
Such lightweight load-time code transformations operate on the string representation of the module, as well as the context to which it is about to be bound, to insert permission-checking code into the library before it is loaded.

\heading{Evaluation}
\sys's evaluation shows that only 7 out of 1044 (0.67\%) unique accesses performed in the tests of 50 libraries were not automatically inferred.
\sys defends against 16 real attacks, only requiring a manual modification of the infered permissions for one attack.
In terms of runtime performance, \sys's static analysis averages \topThousandAnalysisTimePerLib{} per library and its enforcement overhead remains around 1.93\% of the unmodified runtime.

Sections \ref{contracts}--\ref{eval} present our key contributions:

\begin{itemize}
  \item \emph{Permission model and language}:
     a simple yet effective permission model and associated expression language for controlling the functionality available to modules~\sx{contracts},
  \item \emph{Automated permission inference}: 
     a permission inference component that aids developers by analyzing libraries to generate their default permissions~\sx{analysis},
  \item \emph{Privilege reduction quantification}:
      a conceptual framework for quantifying privilege reduction, which allows reasoning about the privilege achieved by applying \sys~\sx{pr},
  \item \emph{Backward-compatible runtime enforcement}:
      a series of program transformations applied upon library load to insert permission checks, interposing on interactions with other libraries and core language structures~\sx{core},
  \item \emph{Implementation and evaluation}:
      an open-source implementation~\sx{eval} as an easily pluggable, backward-compatible library, along with an extensive compatibility \sx{q1}, security (\S\ref{q2}--\ref{q3}), and performance \sx{q4} evaluation.
\end{itemize}

\noindent
Other sections include an example illustrating dynamic compromise and how \sys addresses it~\sx{bg}, a discussion of the threat model~\sx{threat-model}, \sys's relation to prior work~\sx{related}. %
We conclude~\sx{conclusion} that \sys's automation and performance characteristics make it 
  an important addition to a developer's modern toolkit, in many circumstances working in tandem with defenses that focus on other threats.

\section{Background and Overview}
\label{bg}

This section uses a server-side JavaScript application to present (1) security issues related to third-party code~\sx{example}, and (2) an overview of how \sys addresses these issues~\sx{overview}.

\begin{figure}[t]
\includegraphics[width=0.45\textwidth, left]{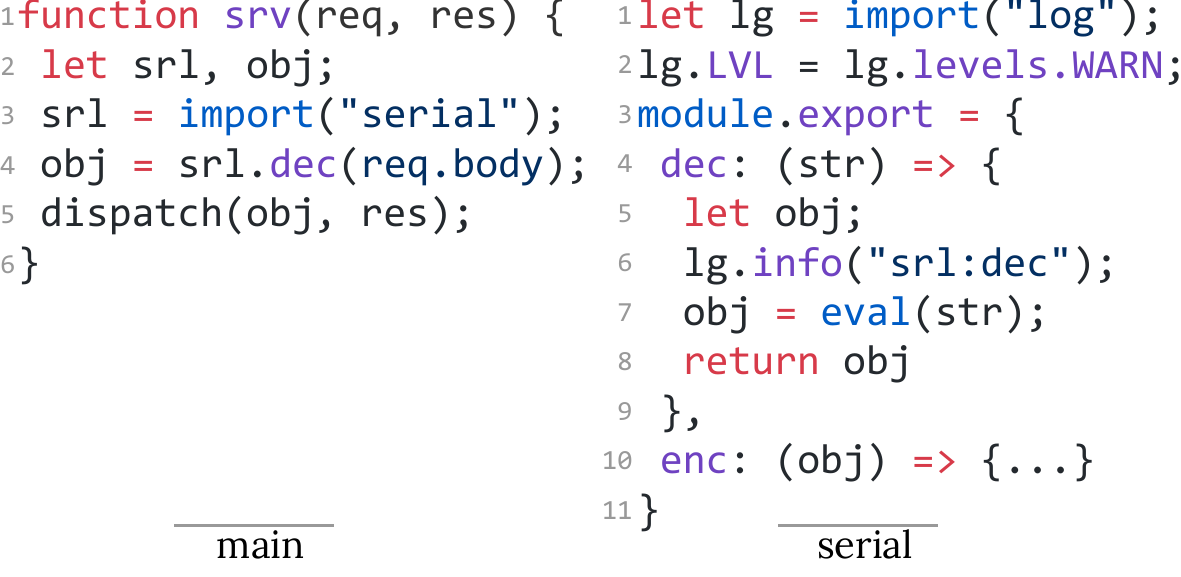}
\caption{
  \textbf{Use of third-party modules.}
  The \ttt{main} module (left) \ctx{import}s off-the-shelf serialization implemented by the \ttt{serial} \emph{third-party} module (right), vulnerable to remote code execution.
}
\label{fig:example}
\end{figure}

\subsection{Example: A De-serialization Library}
\label{example}

Fig.~\ref{fig:overview} presents a simplified schematic of a multi-library program.
Dotted boxes correspond to the context of different third-party modules of varying trust, one of which is used for (de-)serialization.
This module is fed client-generated strings, often without proper sanitization~\cite{synode}, which may enable remote code execution (RCE) attacks.
RCE problems due to serialization have been identified in widely used libraries~\cite{sjs1, sjs2, sjs3}
  as well as high-impact websites, such as PayPal~\cite{paypal} and LinkedIn~\cite{linkedin}.
Injection and insecure de-serialization are respectively ranked number one and eight in OWASP's ten most critical web application security risks~\cite{owasp17}.

For clarity of exposition, Fig.~\ref{fig:example} zooms into only two fragments of the program.
The \ctx{main} module (left) \ctx{import}s off-the-shelf serialization functionality through the \ctx{serial} module, whose \fld{dec} method de-serializes strings using \ctx{eval}, a native language primitive.
The \ctx{serial} module \ctx{import}s \ctx{log} and assigns it to the \ttt{lg} variable.\footnote{
  Naming is important in this paper:
    we differentiate between the module \ctxt{log} and the variable \ttt{lg}.
  \sys tracks permissions at the level of modules, irrespective of the variables they are assigned to.
  To aid the reader, modules, and more broadly, contexts, are typeset in \ctxt{purple sans serif}, fields in \fldt{olive teletype}, and plain variables in \ttt{uncolored teletype} fonts.
}

Although \ctx{serial} is not actively malicious, it is subvertible by attackers at runtime, who can use the input \ttt{str} for several attacks:
(1) overwrite \fld{info}, affecting all (already loaded) uses of \ctx{log}.\fld{info} across the entire program;
(2) inspect or rewrite top-level \ctx{Object} handlers, built-in libraries, such as \ctx{crypto}, and the \ctx{cache} of loaded modules;
(3) access global or pseudo-global variables such as \ctx{process} to reach into environment variables; and
(4) load other modules, such as \ctx{fs} and \ctx{net}, to exfiltrate files over the network.

\subsection{Overview: Applying \sys on Serial}
\label{overview}

Our work address these security problems by intercepting applications at the boundaries between libraries.
\sys's permission model annotates functionality that is not part of the current library with \rwx permissions.
A part of this functionality comes from imported libraries;
  for example, among other permissions, \ctx{serial} needs to be able to execute \fld{info} from module \ctx{log}---\ie \ctx{log}.\fld{info} needs \X.
Another part of this functionality comes directly from the programming language;
  for example, \ctx{serial} clearly needs \X for \ctx{import} and \ctx{eval}.
It is not \ctx{serial} that provides this functionality, but rather the language and its runtime environment.

\begin{figure}[t]
\includegraphics[width=0.25\textwidth, left]{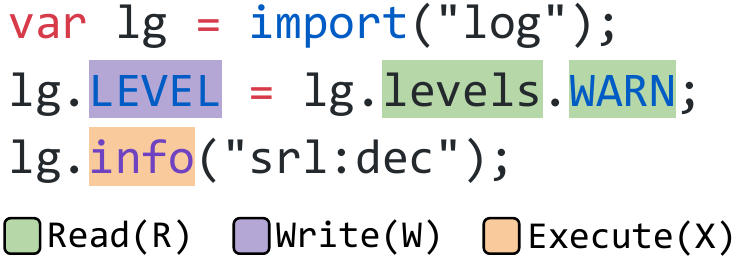}
\caption{
  \textbf{\new Inference.}
  \new uses static analysis to infer module permissions. In this code snippet,
  the different fields of the \ctxt{log} module are given
  distinct permissions based on how they are used.
}
\label{fig:permissions}
\end{figure}

These three permissions are a part of the total nine required for \ctx{serial}'s normal operation. 
To aid developers in identifying the remaining permissions, \sys comes with a static inference component that analyzes how libraries use the available names.
Fig.~\ref{fig:permissions} shows a small example of this analysis:
  \fld{levels} and \fld{WARN} are read, thus are annotated as \R;
  \fld{LEVEL} is written, thus \W; and
  \fld{info} is executed, thus \X.
The analysis also infers permissions for \ctx{import}, \ctx{eval}, \ctx{module}, and \fld{exports}.
Names that do not show up---even if they are built-in language objects---get no permissions.

After extracting all necessary permissions, the developer can start the program using \sys's runtime enforcement component.
\sys shadows all variable names that cross a boundary with variables that point to modified values.
When accessing a modified value, \sys checks the permissions before forwarding access to the original value.
If a module does not have permission to access a value, the approach throws a special exception that helps the developer in diagnosing its cause.

The attacks described at the end of the previous section~\sx{example} are now impossible:
(1) overwriting \fld{info} from \ctx{serial} will throw a \W violation, 
(2) inspecting \ctx{Object} handlers, built-in libraries, such as \ctx{crypto}, and the \ctx{cache} of loaded modules will all throw \R violations, and
(3) accessing global or pseudo-global variables, such as \ctx{process}, to reach into the environment will also throw \R violations.
Shielding against (4) module loading depends on a refinement on the base \rwx model~\sx{contracts}; 
  this refinement does not make automated static inference intractable, as \sys just notes that \ctx{import} is only executable with argument \ttt{log}.

\section{Threat Model}
\label{threat-model}

\sys focuses on the dynamic compromise of possibly buggy or vulnerable libraries, \ie ones that are not actively malicious or hiding their intentions.
Such libraries are subvertible by attackers providing payloads through web interfaces, programmatic APIs, or shared interfaces.
Of particular interest are libraries that offer some form of object de-serialization or runtime code evaluation, where attackers can inject or execute arbitrary code by passing carefully constructed payloads.
This is because these libraries implement their features using runtime interpretation, and thus are subvertible when receiving potentially attacker-controlled inputs (see \S\ref{eval}).

\heading{Focus:}
\sys focuses on the confidentiality (\eg read global state, load other libraries, exfiltrate data) and integrity (\eg write global state or tamper with the library cache) of data and code.
These concerns extend to the broader environment within which a program is executing---including environment variables, the file system, or the network.
Such ambient over-privilege stems from: 
  (1) common features in programming languages (\eg call stack inspection, reflection capabilities, monkey patching);
	(2) unusual language features or deficiencies (\eg in JavaScript: default-is-global, mutability attacks);
	(3) implementation concerns (\eg library cache, import capabilities); and
	(4) authority confusions, where all parts of a program have equal access rights (\eg read \ttt{process.env} or \ttt{process.args}, write to the file-system or network). 

\new aims to mitigate the aforementioned attack vectors by allowing users to control ambient over-privilege at a fine granularity---that of individual names, functions, and fields in the scope of a module.
\sys does not limit itself to the interface exposed explicitly by a module, but rather captures the full observable behavior around it:
  any name that resolves to a value defined outside the module is protected---including language features, built-in names, and environment constructs accessible programmatically.%

\heading{Assumptions:}
\sys's static analysis is assumed to be performed prior to execution, otherwise a malicious library can rewrite the code of a benign library upon load.
For the same reason, \sys's runtime enforcement component is assumed to be loaded prior to any other library.
\new places trust in the language runtime and built-in modules, such as \ttt{fs}:
  a minimum of trusted functionality is needed from the module system to locate and load permissions.

\new does not consider native libraries written in lower-level languages, such as C/C++, or available as binaries.
These libraries are out of scope for two reasons.
First, they cannot be analyzed by \sys's static analysis, which operates on source code.
Second, they can bypass \sys's language-based runtime protection, which depends on memory safety.
Any operation violating memory safety could access arbitrary memory regions.
\sys also does not consider availability, denial-of-service, and side-channel attacks.

\section{Permission Model and Language}
\label{contracts}

\begin{figure}[t]
\centering
\[
\begin{array}{rll}
  s, m& \in \mathsf{String} \\
  r :=& \ttt{R} ~\vert~ \epsilon & \mathsf{ReadPerm} \\
  w :=& \ttt{W} ~\vert~ \epsilon & \mathsf{WritePerm} \\
  x :=& \ttt{X} ~\vert~ \epsilon & \mathsf{ExecPerm} \\
  i :=& \ttt{I} ~\vert~ \epsilon & \mathsf{ImpoPerm} \\
  \mu :=& [r ~w ~x ~i] & \mathsf{Mode} \\
  f :=&  f.s ~\vert~ \str.f ~\vert~ f.\str ~\vert~ s & \mathsf{ObjPath}  \\
  p :=& f \mathbf{:}~\mu ~\vert~ f\mathbf{:}~\mu \mathbf{,}~p &\mathsf{ModPermSet} \\
  \omega :=& m : \mathbf{\{} p \mathbf{\}} ~\vert~ m : \mathbf{\{} p \mathbf{\}}, \omega &\mathsf{FullPermSet} \\
\end{array}
\]
    \caption{
  \textbf{\sys's permission DSL.}
  The DSL captures the set of annotations used for specifying permissions across libraries~\cf{contracts}.
}
\vspace{-10pt}
\label{fig:dsl}
\end{figure}

\new's goal is to reduce the privilege that libraries possess.
At the core of our approach is the ability to explicitly specify a subset of the rights granted to modules by default upon import.
This specification is expressible per-library using a domain-specific language (DSL, Fig.~\ref{fig:dsl}) that focuses on read (\R), write (\W), execute (\X), and import (\I) permissions.

\heading{Core Permission Model}
The core of \sys's permission model and associated DSL is a per-library permission set:
  $\mathsf{ModPermSet}$ maps names accessible within the library context to a $\mathsf{Mode}$, \ie a set of access rights encoded as \rwx permissions.
Ignoring \str-constructs for now, names represent access paths within the object graph reachable from within the scope of the library---\eg \ctx{String}.\fld{toUpperCase}.
Access paths start from a few different points that can be grouped into two broad classes.
The first class contains a closed set of known root points that are provided by the language, summarized in the first four rows of Tab.~\ref{tab:vars}.
These names are available by default through (and shared with) the library's outer context, \ie resolving to a scope outside that of a library and pervasively accessible from any point in the code.
Examples include
  top-level objects and functions, such as \ctx{process}.\fld{args} and \ctx{eval},
  functions to perform I/O, such as \ctx{console}.\fld{log},
  and the library-\ctx{import} ability \emph{itself}.

The second class contains access paths that start from explicitly importing a new library into the current scope.
Such an import results in multiple names available through the imported library's (equivalent of) \ctx{export} statement.
Examples of such paths from Fig.~\ref{fig:example} include \ctx{log}.\fld{info} and \ctx{srl}.\fld{dec}~\sx{example}.

\sys's model can thus be thought as an object-path protection service:
  access rights are expressed as permissions associated with a path from the program's context roots to the field currently accessed.
Values created within the scope of a library or a function are \emph{not} part of this model:
  \new does not allow specifying or enforcing access restrictions on, say, arbitrary objects or function return values.

\heading{Semantics:}
The semantics behind the core set of permissions can be summarized as follows:
\begin{itemize}
  \item 
  A read permission (\R) grants clients the ability to read a value, including assigning it to variables and passing it around to other modules.
  \item
  A write permission (\W) grants clients the ability to modify a value, and includes the ability to delete it.
  The modification will be visible by all modules that have read permissions over the original value.
  \item
  An execute permission (\X)  grants clients the ability to execute a value, provided that it points to an executable language construct, such as a function or a method.
  It includes the ability to invoke the value as a constructor (typically prefixed by \ttt{new}).
\end{itemize}

\noindent
\rwx permissions are loosely based on the Unix permission model, with a few key differences.
Reading a field of a composite value \ttt{x.f} requires \ttt{R} permissions on the value \ttt{x} \emph{and} the field \ttt{f}---that is, an \ttt{R} permission allows only a single de-reference.
Reading or copying a function only requires an \ttt{R}-permission, but performing introspection requires \ttt{X} permissions over its subfields due to introspection facilities being provided by auxiliary methods (\eg \ttt{toString} method).
A \ttt{W} permission on the base pointer allows discarding the entire object.
While a base write may look like it bypasses all permissions, modules holding pointers to fields of the original value will not see any changes.

\heading{Example:}
To illustrate the base permission model on the de-serialization example~\sx{fig:example}, consider \ttt{main}'s permissions:

\begin{lstlisting}[language=js,mathescape,upquote=true]
main: 
  import: RX
  import("serial").dec: RX
\end{lstlisting}

\noindent
The set of permissions for \ttt{serial} is more interesting:
\begin{lstlisting}[language=js,mathescape,upquote=true]
serial:
  eval: RX
  import: RX
  module: R
  module.exports: W
  import("log").levels: R
  import("log").levels.WARN: R
  import("log").info: RX
  import("log").LVL: W

\end{lstlisting}

\begin{table}%
\center
\footnotesize
\caption{
  \textbf{Object path roots.}
  Objects resolving to a scope outside that of a module can be reached through a few starting points:
  (1) core built-in objects,
  (2) the standard library,
  (3) implementation-specific objects,
  (4) module-locals, and
  (5) global variables.
}
\vspace{-1mm}
\begin{tabular}{ll}
  \toprule
  Root Context        & Example Names               \\
  \midrule
  \ctxt{es}            & \ttt{Math}, \ttt{Number}, \ttt{String}, \ttt{JSON}, \ttt{Reflect}, ...    \\
  \ctxt{node}          & \ttt{Buffer}, \ttt{process}, \ttt{console},  \ttt{setImmediate}, ...      \\
  \ctxt{lib-local}     & \ttt{exports}, \ttt{module.exports}, \ttt{\_\_dirname}, ...               \\ %
  \ctxt{globs}         & \ttt{GLOBAL}, \ttt{global}, \ttt{Window}                                  \\
  \ctxt{import}        & \ttt{import($lib$)}, \\
  \bottomrule
\end{tabular}
\label{tab:vars}
\end{table}

\begin{table*}[t]
	\small
  \center
	\caption{
    \textbf{\sys's analysis updates.}
    Updates performed by the static analysis when visiting specific kinds of statements.
    \vspace{-1mm}
  }
	\begin{tabular}{@{}lll@{}}
		\toprule
		Kind of statement & Updates & Example \\
		\midrule
		Assignment $lhs = rhs$ at location $l$: \\
		\hspace{.8em} For each $a \in \mathit{getAPIs}(lhs)$ & Add $(a, \ttt{W})$ to contract $C$ & \ttt{someModule.foo = 5} \\
		\hspace{.8em} For each $a \in \mathit{getAPIs}(rhs)$ & Add $(a, \ttt{R})$ to contract $C$  & \ttt{x = import("someModule")}\\
		& Add $lhs~at~l \mapsto a$ to $\mathit{DefToAPI}$ \\
		Call of function $f$: \\
		\hspace{.8em} For each $a \in \mathit{getAPIs}(f)$ & Add $(a, \ttt{X})$ to contract $C$ & \ttt{someModule.foo()} \\
		Any other statement that contains a reference $r$: \\
		\hspace{.8em} For each $a \in \mathit{getAPIs}(r)$ & Add $(a, \ttt{R})$ to contract $C$  & \ttt{foo(someModule.bar)}\\			
		\bottomrule
	\end{tabular}
	\label{tab:staticAnalysis}
\end{table*}

\heading{Importing}
A simple \X permission to the built-in \ttt{import} function gives libraries too much power.
Thus, \sys needs to allow specifying which imports are permitted from a library.

This is achieved through an additional \I permission. 
This permission is provided to an $\mathsf{ObjPath}$ that explicitly specifies the absolute file-system path of a library.\footnote{
  For portability, \sys prefixes records with a \ttt{\_\_CWD\_\_} variable that can be instantiated to different values across environments.
}
Using the absolute file-system path is a conscious decision:
  the same library name imported from different locations of a program may resolve to different libraries residing in different parts of the file system and possibly corresponding to different versions.

Using a separate permission \I provides additional flexibility by distinguishing from \R.
Libraries are often imported by a part of the program only for their side-effects (\ie not for their interface).
In these cases, their fields should not necessarily be accessible by client code.
Typical examples include singleton configuration objects and stateful servers.

\heading{Wildcards:}
It is often helpful or necessary to provide a single permission mode to \emph{all} possible matches of a segment within a path.
To achieve this, \sys offers \textit{wildcards}:
  \str.\fld{f} assigns a mode to all fields named \fld{f} reachable from any object, and \ctx{o}.\str assigns a mode to all fields of an object (or path) \ctx{o}.
These forms may also be combined, as in \ctx{o}.\str.\fld{f}.

Wildcards have many practical uses.
The primary use case is when fields or objects are altered through runtime meta-programming.
In such cases, the fields are not necessarily accessible from a single static name and might depend on dynamic information.
Often, these fields (not just the paths) are constructed at runtime, which means that they are not available for introspection by \sys at library-load time.

\section{Permission Inference}
\label{analysis}

To aid users in expressing permissions, \sys provides an analysis that automatically infers permissions for a library and its dependencies.

\heading{Static Analysis}
The core of the analysis is an intra-procedural, flow-sensitive forward data flow static analysis.
The analysis is conservative in the sense that if it infers a permission, then there exists a path in the analyzed module that uses the permission.
In contrast, the analysis may miss required permissions, and we evaluate how often this happens in practice.

The analysis visits each statement of a module by traversing a control flow graph of each function.
During these visits, it updates two data structures.
First, it updates the set $C$ of (API, permission) pairs that eventually will be reported as the inferred permission set.
The set $C$ grows monotonically during the entire analysis, \ie the analysis adds permissions until all uses of third-party code have been analyzed.
Second, the analysis updates a map $\mathit{DefToAPI}$, which maps definitions of variables and properties to the fully qualified API that the variable or property points to after the definition.
For example, when visiting a definition \ttt{x = import("foo").bar}, the analysis updates $\mathit{DefToAPI}$ by mapping the definition of \ttt{x} to ``foo.bar''.
The $\mathit{DefToAPI}$ map is a helper data structure discarded when the analysis completes analyzing a function.

Table~\ref{tab:staticAnalysis} summarizes how the analysis updates $C$ and $\mathit{DefToAPI}$ when visiting specific kinds of statements.
The updates to $C$ reflect the way that the analyzed module uses library-external names. %
Specifically, whenever a module reads, writes, or executes an API $a$, then the analysis adds to $C$ a permission $(a, \ttt{R})$, $(a, \ttt{W})$, or $(a, \ttt{X})$, respectively.
The updates to $\mathit{DefToAPI}$ propagate the information about which APIs a variable or property points to.
For example, suppose that the analysis knows that variable \ttt{a} points to a module ``foo'' just before a statement \ttt{b = a.bar}; then it will update $\mathit{DefToAPI}$ with the fact that the definition of \ttt{b} now points to ``foo.bar''.

While traversing the control flow graph, the analysis performs the updates in Table~\ref{tab:staticAnalysis} for every statement.
On control flow branches, it propagates the current state along both branches.
When the control flow merges again, then the analysis computes the union of the $C$ sets and the union of the $\mathit{DefToAPI}$ maps of both incoming branches.
\sys handles loops by unrolling each loop once, which is sufficient in practice for analyzing uses of third-party code, because loops typically do not re-assign references to third-party APIs.

\heading{Qualified Interfaces}
The update functions in Table~\ref{tab:staticAnalysis} rely on a helper function $\mathit{getAPIs}$.
Given a reference, \eg a variable or property access, this function returns the set of fully qualified APIs that the reference may point to.
For example, after the statement \ttt{obj.x = import("foo").bar}, $\mathit{getAPIs}(obj.x)$ will return the set $\{``foo.bar''\}$. 
When queried with a variable that does not point to any API, $\mathit{getAPIs}$ simply returns the empty set.
Algorithm~\ref{alg:getAPIs} presents the $\mathit{getAPIs}$ function in more detail.
We distinguish four cases, based on the kind of reference given to the function.
Given a direct import of a module, $\mathit{getAPIs}$ simply returns the name of the module.
Given a variable, the function queries pre-computed reaching-definitions information (see below) to obtain possible definitions of the variable, and then looks up the APIs these variables point to in $\mathit{DefToAPI}$.
Given a property access, \eg \ttt{x.y}, the function recursively calls itself with the reference to the base object, \eg \ttt{x}, and then concatenates the returned APIs with the property name, \eg ``y''.
Finally, for any other kind of reference, $\mathit{getAPIs}$ returns an empty set.
This case includes cases that we cannot handle with an intra-procedural analysis, \eg return values of function calls.
In practice, these cases are negligible, because real-world code rarely passes 
around references to third-party APIs via function calls.
We therefore have chosen an intra-procedural analysis, which ensures that the 
static contract inference scales well to large code-bases.

\begin{algorithm}[tb]
	\small
	\KwData{Reference $r$}
	\KwResult{Set of APIs that $r$ may point to}

	\If{$r$ {\normalfont is an import of module ``m''}}{
		\textbf{return} \{ ``m'' \}
    }

	\If{$r$ {\normalfont is a variable}}{
		$A \leftarrow \emptyset$
		
		$\mathit{defs} \leftarrow$ get reaching definitions of $r$
		
		\For{\textbf{each} $d$ \textbf{in} $\mathit{defs}$}{
			$A \leftarrow A \cup \mathit{DefToAPI}(d)$
		}
	
		\textbf{return} $A$
	}

	\If{$r$ {\normalfont is a property access} $\mathit{base}.\mathit{prop}$}{
		$A_{\mathit{base}} \leftarrow \mathit{getAPIs}(\mathit{base})$
	
		\textbf{return} $\{ a + "." + \mathit{prop} ~|~ a \in A_{\mathit{base}}\}$
	}
  
  	\textbf{return} $\emptyset$
  	
	\caption{Helper function $\mathit{getAPIs}$.}
	\label{alg:getAPIs}
\end{algorithm}

To find the APIs a variable may point to, Algorithm~\ref{alg:getAPIs} gets the reaching definitions of the variable.
This part of the analysis builds upon a standard intra-procedural may-reach 
definitions analysis, which \sys pre-computes for all functions in the 
module. To handle nested scopes, \eg due to nested function definitions, \sys
builds a stack of definition-use maps, where each scope has an associated set of 
definition-use pairs.
To find the reaching definitions of a variable, the analysis first queries the inner-most scope, and then queries the surrounding scopes until the reaching definitions are found.
To handle built-in APIs of JavaScript, \eg \ttt{console.log}, \sys creates an artificial outer-most scope that contains the built-in APIs available in the global scope.

Returning to the running example in Figure~\ref{fig:example}.
For \ctx{main}, the static analysis results in the following permission set:
$$\{(``serial'', \ttt{R}), (``serial.dec'', \ttt{R}), (``serial.dec'', \ttt{X})\}$$
As illustrated by the example, the inferred contract allows the intended behavior of the module, but prevents any other, unintended uses of third-party APIs.
Our evaluation shows that the static analysis is effective also for larger, 
real-world modules~\sx{q2}.

\heading{Limitations}
In line with \sys's design goal of being conservative in granting permissions, the analysis infers a permission only if there exists a path that uses the permission.
In contrast, the analysis may miss permissions that a module requires.
For example, missed permissions may results from code that passed a reference to a module across functions:
\begin{lstlisting}[language=js,mathescape,upquote=true]
x = require("foo");
bar(x);
\end{lstlisting}
In this example, the analysis misses any permissions on ``foo'' that are required by \ttt{bar}.
Tracking object references across function boundaries would require an inter-procedural analysis, which is difficult to scale to a module and its potentially large number of transitive dependencies~\cite{npmstudy:19}.
Another example of potentially missed permissions is code that dynamically computes property names:
\begin{lstlisting}[language=js,mathescape,upquote=true]
x = require("foo");
p = "ba" + "r";
x[p] = ...;
\end{lstlisting}
In this example, the analysis misses the \ttt{W} permission for ``foo.bar''.
Tracking such dynamically computed property names is known to be a hard problem in static JavaScript analysis~\cite{Sridharan2012}.

\heading{Import-time Permission Analysis}
To reduce the number of required permissions not inferred by its static analysis, \sys adds a short phase of dynamic import-time analysis.
This analysis is performed by simply importing a library---but without invoking the library's interface---and only adds to the number of permissions inferred by the analysis described earlier.
The underlying insight is that many libraries simply wrap or re-export existing interfaces using dynamic meta-programming.
The following code snippet demonstrates a simple but common pattern:
\begin{lstlisting}[language=js,mathescape,upquote=true]
for (let k in fs) {
  module.exports[k] = fs[k];
}
\end{lstlisting}
Inferring such meta-programming permissions statically poses a challenge due to the limitations discussed earlier, and thus simply loading the library should be enough to enable a more complete view into the library's behavior.
Import-time analysis does not depend on the existence of library tests or any consuming code, as it does not call any library interfaces.

\section{Quantifying Privilege Reduction}
\label{pr}

Any policy---whether automated or manual---on existing programs aims at striking a balance between compatibility and security:
  an ideal policy would allow only the necessary accesses but no more.
Unfortunately, statically inferring such an ideal policy in the context of a dynamic language---even with the addition of import-time permissions---is impossible, due to the language's very nature.
However, some analyses are better than others, \ie they infer policies with fewer accesses, even if they do not infer the ideal policy.
To be able to quantitatively evaluate the security benefits offered by such analyses, we propose a novel privilege reduction metric. 

\heading{Privilege Reduction}
Informally, the single-library privilege reduction is calculated as the ratio of disallowed permissions over the full set of permissions available by default within the lexical scope of the library.
The default permission set is calculated by statically expanding all names available in scope; 
  the disallowed set of permissions is calculated by subtracting the allowed permissions from the default permission set.
Single-library privilege reductions across the full dependency tree are then combined into a single reduction metric for the program's current decomposition.
The following paragraphs explain the details.

The quantitative definition of privilege reduction depends on a few key notions.
The first notion is a set of \emph{target critical resources} $M_t$ that are used to restrict a library's (or program's) access.
The second notion is the set of \emph{subjects} $M_s$ that can access these resources through specific means (the permissions).
\sys's quantitative definition of privilege reduction is  general in that it can be applied to any scenario with a finite number of subjects and critical resources.

\heading{Informal Development}
Before formalizing privilege reduction, we use the de-serialization example to build an intuition.
Let's assume that from the two modules presented in Fig.~\ref{fig:example}, we are only interested in quantifying \ctx{main}'s prvilege;
  thus, $M_s = \{ \ctx{main} \}$. 
As implied earlier~\sx{contracts}, the set of critical resources contains many paths available to \ctx{main}.
For simplicity, we now assume it only contains \ctx{globals}, \ctx{fs}, and \ctx{import};
  thus,  $M_t = \{ \ctx{globals}, \ctx{fs}, \ctx{import} \}$.
Module \ctx{main} needs an \X permission on \ctx{import} to be able to load \ctx{serial}, and an \X permission on \ctx{serial}.\fld{dec} to be able to call the \ttt{dec} function.
With this simple configuration, \sys disallows all accesses except for
   $P(M_s, M_t) = \{ \langle \ctx{import}, X \rangle, \langle \ctx{serial}.\fld{dec}, X \rangle \}$.

\sys's goal is to quantify this privilege with respect to the default permissions.
If \ctx{main} was executed without additional protection, its privilege would be
  $P_{base}(M_s, M_t) = \{ \langle \ctx{globals}.\fld{*}, RWX \rangle, \langle \ctx{fs}.\fld{read}, RWX \rangle, ... \}$.

\heading{Formal Development}
More formally, by default at runtime any module has complete privilege on all exports of any other module.
Thus, for any modules $m_1, m_2$ the baseline privilege that $m_1$ has on $m_2$ is:
\[
  P_{base}(m_1, m_2) = \{ \langle a, \rho \rangle | a \in API_{m_2},
  \rho \in P \}
\]
where $\rho \in \mathcal{P}$ is a set of orthogonal permissions on a resource, which for \sys is  $\mathcal{P} = \{ R, W, X \}$.
Name $a$ can be any field that lies arbitrarily deeply within the values exported by another module. %

\sys reduces privilege by disallowing all but the white-listed permissions at module boundaries:
\[
P_S(m_1, m_2) = \{ \langle a, \rho \rangle | a \in API_{m_2}, S \text{
  gives } m_1 \  \rho \text{ on } a\}
\]
To calculate the privilege reduction across a program that contains several different modules, we lift the privilege definition to a set of subject and target modules:
\[
P(M_s, M_t) = \bigcup_{\substack{m_1 \in M_s \\ m_2 \in M_t}} P(m_1, m_2) 
\]

Based on this, we can define privilege reduction, a metric of the permissions restricted by a privilege-reducing system $S$, such as \sys:
\[
PR(M_s, M_t) = \frac{|P_{base}(M_s, M_t)|}{|P_{S}(M_s, M_t)|}
\]
A higher reduction factor implies a smaller attack surface since the subjects are given privilege to a smaller portion of the available resources.
$P_{base}$ is an under-approximation of base privileges, as a source module can in principle import and use any other malicious module that is installed in the execution environment.
Consequently, the measured privilege reduction is actually a lower bound of the privilege reduction that \sys achieves in practice.

\begin{figure*}[t]
\centering
\includegraphics[width=0.98\textwidth]{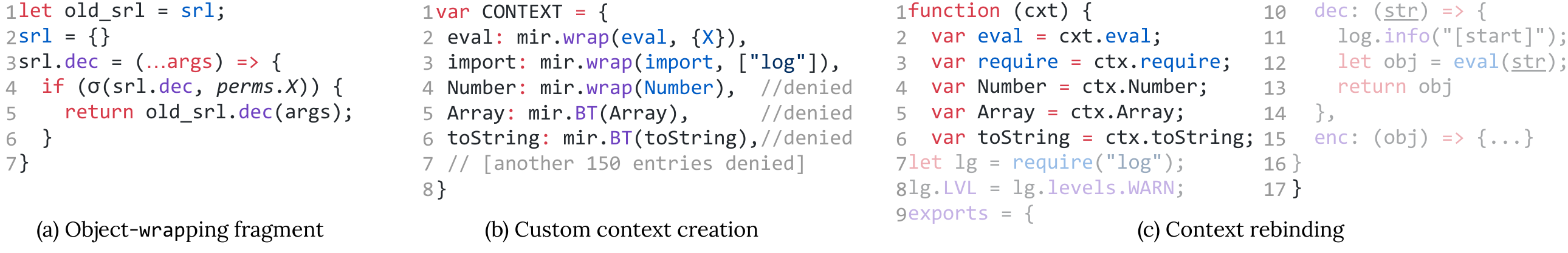}
\caption{
  \textbf{\sys's runtime enforcement transformations.}
  \sys's basic \ttt{wrap}ping traverses objects and wraps fields with inline monitors (a, line 4).
  A new modified context is created by \ttt{wrap}ping all values available in a module's original context (b).
  The modified context is bound to the module by enclosing the module source (half-visible code fragment) in a function that redefines all non-local variable names as function-local ones, pointing to values from the modified context~\cf{core}.
\vspace{-12pt}
}
\label{fig:enforcement}
\end{figure*}

\heading{Permission importance}
An important observation is that different permissions $\rho \in R$ may not be equally important.
In principle, \X may be considered more dangerous than \W, which in turn may be considered more dangerous than \R.
Similarly, in terms of ``base'' APIs, \ctx{console}.\fld{log} may be considered less important than \ctx{fs}.\fld{write}.
For these reasons, \sys's model abstracts these values as coefficients; 
  the same coefficients should be used for meaningful comparisons across different configurations.
In this paper we set all coefficients to the value ``1'', considering all permissions to be equally important.

\heading{Transitive Permissions}
Fig.~\ref{fig:example}'s \ctx{main} is not allowed to directly call \ctx{eval};
  however, it can call \ctx{eval} indirectly by executing \ctx{serial}.\fld{dec}.
This is a limitation of attempting to calculate the privilege reduction dynamically within an environment that specifies opaque abstraction boundaries, such as libraries.
Accurately quantifying such transitive privilege requires tracking transitive calls across such boundaries, which requires heavyweight information flow analysis.
\sys's privilege reduction quantification does not attempt such an analysis to keep runtime overheads low.
As a result, \sys's estimate is necessarily conservative---\ie \sys reports a lower number than the one achieved in practice.

\section{Runtime Permission Enforcement}
\label{core}

During program execution, \sys's runtime component enforces the chosen permissions---automatically inferred, developer-provided, or a combination thereof.
\sys's enforcement is enabled by the fact that JavaScript features a module-import mechanism that loads code at runtime as a string.
Lightweight load-time code transformations operate on the string representation of each module, as well as the context to which it is about to be bound, to insert enforcement-specific wrappers into the module before it is loaded.

\sys's transformations can be grouped into four phases.
The first phase simply modifies \ctx{import} so that calls yield into \new rather than the built-in locate-and-load mechanism.
For each module, the second phase creates a fresh copy of the runtime context---\ie all the name-value mappings that are available to the module by default.
The third phase binds the modified context with the module, using a source-to-source transformation that re-defines names in the context as library-local ones and assigns to them the modified values.
After interpreting the module, the fourth phase further transforms the module's interface so that its client can only access the names---\eg methods, fields---it is allowed to access.

\begin{table*}[t]
\center
\footnotesize
\setlength\tabcolsep{2pt}
\caption{
  \textbf{Invalid accesses and reasons for incompatibility.}
  For libraries that showed some incompatibility, the table shows the number of inferred permissions (col. 2, 3),  the number of invalid permissions (col. 4, 5), and the primary reasons for each incompatibility (col. 6).
   Permissions include two configurations---base static analysis (col. 2, 4) and augmented with import-time inference (col. 3, 5). 
}
\begin{tabular*}{\textwidth}{r c cc c cc c l@{\extracolsep{\fill}}}
\toprule
               & & \multicolumn{2}{c}{Inferred}  & &  \multicolumn{2}{c}{Missing}            & &                       \\
                   \cline{2-4}                                    \cline{6-8}
               & & base & +iti                    & ~~ &  base  & +iti & & ~~Reason for incompatibility \\
\midrule
through2       & &  18  & 24                     &    & 2      & 0     & & Gets (inherits) properties through a testing library       \\
set-value      & &  14  & 14                     &    & 3      & 3     & & Dynamic access to \ttt{process.env.npm\_package\_keywords\_15} \\
periods        & &  6   & 20                     &    & 14     & 0     & & Function call in \ttt{export} index      \\
node-slug      & &  9   & 10                     &    & 1      & 0     & & Dynamic property access to \ttt{module.exports.*}   \\
is-generator   & &  7   & 7                      &    & 2      & 2     & & \ttt{fn.constructor} does not see \ttt{fn} as a parameter \\
he             & &  12  & 17                     &    & 6      & 2     & & Access to \ttt{process.env.npm\_package\_keywords\_4, process.env.npm\_package\_keywords\_5}\\
fs-promise     & &  14  & 180                    &    & 135    & 0     & & Dynamically computed across all methods of \ttt{require("fs")} \\
file-size      & &  9   & 10                     &    & 1      & 0     & & Assignment at return: \ttt{return exports = plugin()} \\
\midrule
  total        & & 89   & 282                    &    & 175    & 7     & & 95.5\% Improvement due to import-time inference      \\
\bottomrule
\end{tabular*}
\label{tab:reasons}
\end{table*}

\heading{Base Transform}
These transformations have a common structure that traverses objects recursively---a base transformation \ttt{wrap}, which we review first.
At a high level, \ttt{wrap} takes an object $O$ and a permission set $p$ and returns a new object $O'$.
Every field $f$ of $O$ is wrapped with a method $f'$ defined to enclose the permissions for $f$.
Every $f'$ implements a security monitor---a level of indirection that oversees accesses to the field and ensures that they conform to the permissions corresponding to that field.
At runtime, $f'$ checks $f$'s permission for the current access type:
 if the access is allowed, it forwards the call to $f$;
  otherwise, it throws a special exception, \ttt{AccessControlException}, that contains contextual information for diagnosing root cause---including the type of violation (\eg \R), names of the modules involved, names of accessed functions and objects, and a stack trace.

The result of applying the \ttt{wrap} transformation to the object (returned by) \ttt{serial} is shown in Fig.~\ref{fig:enforcement}a.

\heading{Context Creation}
To prepare a new context to be bound to a library being loaded,
  \sys first creates an auxiliary hash table (Fig.~\ref{fig:enforcement}\emph{b}), mapping names to (newly transformed) values:
  names correspond to implicit modules---globals, language built-ins, module-locals, \etc (Tab.~\ref{tab:vars});
  transformed values are created by traversing individual values in the context using the \ttt{wrap} method to insert permission checks.

User-defined global variables are stored in a well-known location (\ie a map accessible through a global variable named \ttt{global}).
However, traversing the global scope for built-in objects is generally not possible.
To solve this problem, \sys collects such values by resolving well-known names hard-coded in a list.
Using this list, \sys creates a list of pointers to unmodified values upon startup.

Care must be taken with module-local names---\eg the module's absolute filename, its \ttt{export}ed values, and whether the module is invoked as the application's \ttt{main} module:
  each module refers to its own copy of these variables.
Attempting to access them directly from within \sys's scope will fail subtly, as they will end up resolving to module-local values of \sys \emph{itself}---and specifically, the module within \sys applying the transformation.
\sys solves this issue deferring these transformations for the context-binding phase (discussed next). %

Fig.~\ref{fig:enforcement}b shows the creation of \ttt{serial}'s modified context.

\heading{Context Binding}
To bind the code whose context is being transformed with the freshly created context, \sys applies a source-to-source transformation that wraps the module with a function closure.
By enclosing and evaluating a closure, \sys leverages JavaScript's lexical scoping to inject a non-bypassable step in the variable name resolution mechanism.

The closure starts by redefining default-available non-local names as module-local ones, pointing to transformed values that exist in the newly-created context.
It accepts as an argument the customized context and assigns its entries to their respective variable names in a preamble consisting of assignments that execute before the rest of the module.
Module-local variables (a challenge outlined earlier) are assigned the return value of a call to \ttt{wrap}, which will be applied only when the module is evaluated and the module-local value becomes available.
\sys evaluates the resulting closure, invokes it with the custom context as an argument, and applies further \ttt{wrap} transformations to its return value.

The result of such a source-to-source linking of \ttt{serial}'s context is shown in Fig.~\ref{fig:enforcement}c.

\section{Implementation \& Evaluation}
\label{eval}

\begin{figure*}[t]
  \centering
  \includegraphics[width=\textwidth]{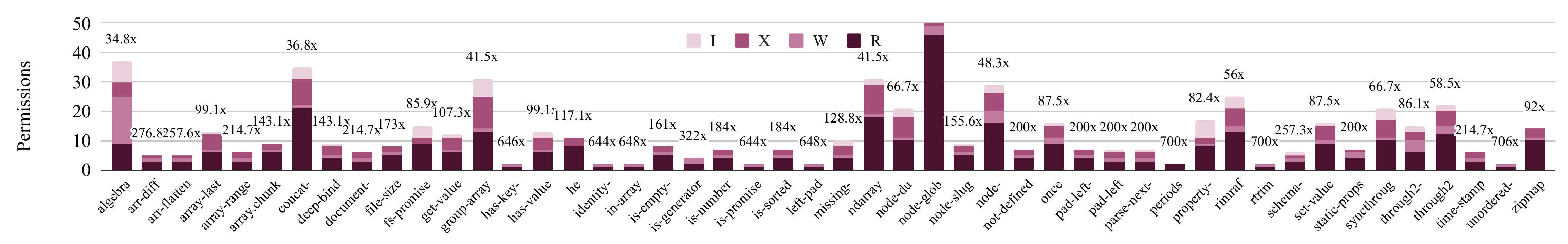}
  \caption{
    \textbf{Inferred Permissions.}
		Number of inferred permissions.
    Stacked bars show the number of \R, \W, \X, and \I permissions;
      the number above the bar shows the privilege reduction over their default permissions~\cf{q2}.
  }
\label{fig:reduction}
\end{figure*}

\noindent
We built a prototype of \sys for JavaScript, targeting the Node.js ecosystem, and available via \ttt{npm i -g @blindorg/mir}.
This command makes a few components available.
\sys's static analysis component~\sx{analysis} becomes available as \ttt{mir-sa}, implemented as a compiler pass in the Google Closure Compiler~\cite{gcci} in about 2.1 KLoC.
\sys's runtime enforcement component~\sx{core} is available as \ttt{mir-da}, implemented in about 2.8 KLoC of JavaScript. 
It supports the configuration of several parameters---for example, \ttt{depth} is the number of levels \sys augments, starting from the root of the top objects (Tab.~\ref{tab:vars}).
Permissions~\sx{contracts} exchanged between \ttt{mir-sa} and \ttt{mir-da} are specified as auxiliary JSON files.
To evaluate \sys, we investigate the following questions:

\begin{itemize}

  \item \textbf{Q1} How compatible is \sys with existing code---\ie what is the danger of breaking legacy programs?
    Only 7 out of 1044 (0.67\%) unique accesses performed in the tests of 50 libraries were not inferred by \sys~\sx{q1}.
  \item \textbf{Q2} How effective is \sys in theory, in terms of privilege reduction?
    \sys provides 15.6--706$\times$ (avg: 224.5$\times$) reduction in terms of permissions over the default execution~\sx{q2}.
  \item \textbf{Q3} How effective is \sys in practice, against \emph{in vivo} and \emph{in vitro} attacks?
    Applied on (1) 16 real attacks in real packages, and (2) several hundred crafted attacks against a highly-vulnerable module, \sys mitigates all attacks that fall under its threat model.
    In all cases except one, \sys applied default permissions inferred automatically~\sx{q3}.
  \item \textbf{Q4} How efficient and scalable are \sys's inference and enforcement components?
    \sys's static analysis averages \topThousandAnalysisTimePerLib{} per library, comparably to a widely-used linter, and its runtime overhead averages about 1.93\%~\sx{q4}. %
\end{itemize}

\noindent
The remainder of the evaluation section presents the supporting evidence and details underlying these findings.

\heading{Libraries and workloads}
To address Q1, Q2, and Q4 we focus on a set of 50 packages from a curated list~\cite{awesome}
that (i) provide comprehensive tests suites executable via \ttt{npm test} and (ii) are small enough to enable us to inspect their behavior, and in particular, their use of other packages.
Criterion (ii) is crucial for not only reporting quantitative but also qualitative results about the compatibility provided by \sys.
As the 50 libraries are quite popular (averaging 4.8M weekly downloads (total: 227M) and 656 dependees (total: 30K)), 
  development investment in their testing infrastructure has been significant, having a variety of different tests that stress different parts of the library---in some cases tests were $10\times$ the size of the corresponding library.
To answer Q3, we use a different set of benchmarks that contain vulnerable code.
A set of 16 vulnerabilities in real packages that were found \emph{in vivo}~\cite{snyk};
  and a series of attacks against a single module crafted \emph{in vitro} to represent the most vulnerable threat within \sys's threat model.
To further evaluate the scalability of the static analysis in Q4, we also apply the analysis to \topThousandpkgsToConsider{} npm packages, which comprise \topThousandSLOC{} LoC.

\heading{Setup}
Experiments were conducted on a modest server with 4GB of memory and 2 Intel Core2 Duo E8600 CPUs clocked at 3.33GHz, running a Linux kernel version 4.4.0-134.
The JavaScript setup uses Node.js v12.19, bundled with V8 v7.8.279.23, LibUV v1.39.0, and \ttt{npm} version v6.14.8.
\sys's static inference run on Java SE 1.8.0\_251, linked with Google Closure v20200927.
Libraries were tested on their latest versions at the time of this writing.%

\subsection{Compatibility Analysis (Q1)}
\label{q1}

To a large extent, backward-compatibility drives practical adoption of tools such as \sys---if a tool requires significant effort to address compatibility, chances are developers will avoid it despite of any security benefits it provides.
This is especially true for \sys, which follows a static analysis approach to meet its threat model of dynamic compromise, rather than one based on dynamic analysis.

\heading{Big Picture}
\sys applies an average of 346 transformations per library, wrapping a total of 25,609 fields.
The distribution of accesses is bimodal: (1) not all fields are accessed at runtime---on average only 20.88 (6.06\%) of fields are accessed per library; (2) the ones that do get accessed, are accessed multiple times: 794.9 times per field, on average.

\piccaption{
    \textbf{Invalid accesses.}
    Only 3 out of 50 libraries encountered an invalid access, totalling 7 out of 1044 unique accesses (0.67\%) in less than 0.1\% of tests~\cf{q1}.
    \label{fig:invalid}
    }
\parpic[r][t]{%
  \begin{minipage}{34mm}
  \includegraphics[width=0.85\columnwidth]{\detokenize{./figs/mir_invalid_accesses}}
  \end{minipage}
}
Fig.~\ref{fig:invalid} shows a histogram of the number of packages and the number of unique invalid accesses, \ie the number of benign accesses that happen in the tests but were not found by \sys's inference component.
From the 50 libraries on which we applied \sys, only 3 showed some divergence between statically inferred permissions and runtime behavior. 
All of these 3 libraries diverged on 3 accesses or less.

These libraries were tested under multiple workloads each triggering different accesses.
Counting with respect to tests, rather than libraries, about 99.9\% of all tests performed no invalid accesses.
From 1044 total unique accesses, only 7 (0.67\%) were invalid;
  if repetitions are taken into account, 12 out of 16,598 (0.07\%) were
  invalid.

\heading{Incompatibility}
Tab.~\ref{tab:reasons} zooms into modules that show some degree of incompatibility between inferred and required permissions.
It shows the number of inferred permissions (col. 2, 3),  the number of invalid permissions (col. 4, 5), and the primary reasons for each incompatibility, under two inference configurations:
  base static analysis (col. 2, 4) and augmented with import-time inference (col. 3, 5).
The key take-away is that \sys's import-time inference
is enough to improve results by 95.5\%:
  it brings the total number of non-inferred accesses to seven (from 175, due to static-only inference).
We next focus on a few interesting cases.

Modules \ttt{zipmap} and \ttt{concat-stream} create a local \fld{toString} that they get via the \ctx{Object} prototype.
\sys's runtime enforcement is configured to not wrap the prototype chain by default, missing the \R.
Moreover, the runtime reports  \fld{toString} as global (capturing a \R on \ttt{global} and \W on \fld{toString}), because it is defined via \ttt{var}.
(Changing \ttt{var} to \ttt{let} leads to base incompatibility of 1.)
Modules 
\ttt{static-props}, 
\ttt{set-value}, 
\ttt{fs-promise} 
\ttt{is-generator}, 
and 
\ttt{he} 
compute properties they access dynamically.
Module \ttt{fs-promise} is a good example of this behavior:
 it traverses the built-in \ttt{fs} interface to generate promise-fied wrappers.

These cases highlight a connection between the size of the interface manipulated and the type of incompatibility showing up.
When developers use many or all of the properties of an object, they are likely to access them programmatically via runtime reflection; 
  conversely, if they use reflection, they are likely to be accessing many (or all of the) properties.
In these cases, static analysis is unlikely to extract these behaviors, which is where import-time inference aids.

The appendix discusses the sources of permission incompatibility between static analysis and dynamic enforcement~\sx{two} in more detail.

\begin{table}[b]
\center
\footnotesize
\setlength\tabcolsep{2pt}
\caption{
  \textbf{\emph{In vivo} attacks defended by \sys.}
  The first column contains vulnerable modules $\times$ number of exploits applied to each module.
  The last column shows the number of permissions inferred by \sys~\cf{q3}.
}
\begin{tabular}{r llll c}
\toprule
Module                   &  Version     & Attack                            &  CVE        & CWE    & Perms\\ %
\midrule
Ser/pt                   & $<1.0$       & Arbitrary Code Execution          & 2017-5941   & 502    & 13   \\ %
Nod/ze                   & $0.0.4$      & Arbitrary Code Execution          & 2017-5954   & 502    & 7    \\ %
Ser/js                   & $<3.1$       & Arbitrary Code Injection          & 2020-7660   & 94     & 58   \\ %
Saf/al  $\times4$        & $<0.4$       & Sandbox Escaping                  & (multiple)  & 265    & 10   \\ %
Stat/al  $\times2$       & $<2.0$       & Arbitrary Code Execution          & 2017-16226  & 94     &  9   \\ %
Fast-redact              & $<1.5$       & Arbitrary Code Execution          &     ---     & 94     & 32   \\ %
Mathjs $\times5$         & $<1.5$       & Arbitrary Code Execution          & (multiple)  & 94     & 1    \\ %
Morgan                   & $<1.9$       & Arbitrary Code Injection          & 2018-3784   & 502    & 27   \\ %
\bottomrule
\end{tabular}
\label{tab:sec}
\end{table}

\subsection{Permissions and Privilege Reduction (Q2)}
\label{q2}

In this section we measure the privilege reduction achieved by \sys's static inference.

\heading{Number of Permissions} 
Fig.~\ref{fig:reduction} shows \sys's number of inferred \R, \W, \X, and \I permissions.
Their number ranges between 3--54 (avg: 11.9), spread unevenly between \R (6.85), \W (1.54), \X (3.42), and \I (1.52).
\sys infers a small number of permissions because developers use only a small subset of the APIs provided by libraries and the broader language environment.

\heading{Privilege Reduction}
Fig.~\ref{fig:reduction} also shows the privilege reduction achieved by \sys---\ie the ratio of allowed permissions over a library's default set of permissions~\sx{pr}.
Privilege is reduced by up to three orders of magnitude, ranging between $15.6\times$--$706\times$ (avg: $224.5\times$).
\sys's privilege reduction is high because developers use only a small fraction of the available APIs.

\begin{figure*}[t]
  \centering
  \includegraphics[width=0.99\textwidth]{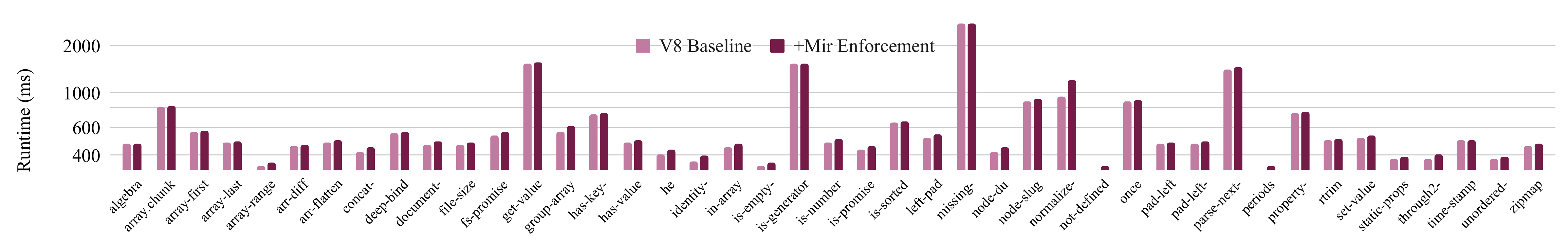}
  \caption{
    \textbf{Performance of \sys's runtime enforcement component.}
    Comparison between vanilla runtime performance with 
      (1) full \rwx with 3-level depth and across all contexts, and 
      (2) 3-level depth across global-only variables.
    On average, +3.3ms per library (1.93\% slowdown) and +3.2ms (1.62\% slowdown)~\cf{q4}.
  }
\label{fig:dperf}
\end{figure*}

\subsection{Practical Security Impact (Q3)}
\label{q3}

This section complements the analysis of \sys's privilege reduction with an investigation of its effectiveness on vulnerable code by applying it to \emph{in vivo} and \emph{in vitro} vulnerabilities.
The former set focuses on packages found in the wild~\cite{snyk}, for which Table~\ref{tab:sec} includes the vulnerable versions and MITRE's common enumeration identifiers~\cite{cve, cwe}.
The latter set focuses on attacks against a carefully crafted scenario that allows completely unrestricted runtime code evaluation, capturing the most dangerous vulnerability within \sys's threat model.
\sys defends against all these vulnerabilities;
  apart from \emph{Mathjs}, no permissions required manual intervention.

\heading{In Vivo}
For these vulnerabilities, we use the proof-of-concept exploit (PoC) attached to the original vulnerability report;
  many of them include multiple CVEs or PoCs, all of which we attempt.
\emph{Ser/pt}, \emph{Nod/ze}, and \emph{Ser/js} all perform some form of unsafe serialization;
  their PoCs either (1) import \ctx{child\_process} to call \ttt{ls} or \ttt{id}, or (2) invoke \ctx{console}.\fld{log}.
As none of these is part of the library's permission set, \sys disallows access to these APIs.

\emph{Saf/al} and \emph{Stat/al} sanitize input prior to calling \ctx{eval}.
The PoC payloads are similar to the ones described earlier, against which \sys protects.
The case of \emph{Stat/al} is interesting because it accepts ASTs rather than strings;
  the PoC passes \ctx{process}.\fld{env} through the Esprima parser making it vulnerable---which \sys solves by disallowing access to \ctx{process}.\fld{env}.

The \emph{Fast-redact} PoC uses a \fld{constructor} to reach into the \ctx{Function} prototype and invoke \fld{open} via \ctx{process} bindings.
\sys allows the first two steps, crucial for compatibility, but blocks the call to both \fld{bindings} and \fld{open}.

The popular \emph{Mathjs} module includes a math evaluator, which the PoC exploits to \ctx{console}.\fld{log} a message.
As \emph{Mathjs} is compiled into a universal-module-definition by webpack (\ie features no imports and thus is a pure function), we provided its permissions manually.
Despite the fact that \emph{Mathjs} does not import any modules, the PoC calls into Node's APIs via \ctx{cos}.\fld{constructor} (\`{a} la \emph{fast-redact}).
While \sys does not block the \ttt{constructor} call, it does block \ctx{process}.\fld{env} and \ctx{fs}.\fld{read} from within \emph{Mathjs}.

For most of these benchmarks, \sys blocks the PoC at multiple levels---\eg even if \emph{Nod/ze}'s import of \ctx{child\_process} was allowed, \sys would still block \fld{exec}.

\pichskip{15pt}%
\piccaption{
    \textbf{\emph{In vitro} attacks prevented by \new}.
    \sys defends against handcrafted remote code execution attacks, a subset of which is shown here.
    \label{tab:vitro}
    }
\parpic[r][t]{%
  \begin{minipage}{34mm}
    \center
    \footnotesize
    \setlength\tabcolsep{2pt}
    \begin{tabular}{l}
    \toprule
    Attack \\
    \midrule
    write ~~\ttt{x}, \ttt{global.x} \\
    read  ~~\ttt{require.cache} \\
    execute \ttt{process.argv} \\
    read \ttt{process.env} \\
    \ttt{fs.read} \ttt{/etc/passwd} \\
    \ttt{chi/ocess.spawn} \\
    call \ttt{Math.log} \\
    read \ttt{os.EOL} \\
    \bottomrule
    \end{tabular}
  \end{minipage}
}
\heading{In Vitro}
The goal of this experiment is to evaluate \sys against the worst-case, most-vulnerable single module that falls under its threat model.
As \sys aims at protecting against dynamic compromise, the ultimate dynamic-compromise vulnerability would be to expose fully unrestricted access to runtime code evaluation.
This is the most general threat that falls within \sys's threat model in the context of a single module, as smaller-scope vulnerabilities for reading and writing fields are subsumed by runtime code execution---and code execution via \ctx{eval} comes with the fewest protections:
  any string passed to \ctx{eval} is evaluated at runtime---effectively exposing a runtime interpreter to the attacker.

To implement this scenario, we construct a library \ttt{e} that only exposes an \ctx{eval} function.
We craft several expressions that attempt to access sensitive interfaces---over 800 attacks amounting to $>$11K possible accesses targeting built-in JavaScript, Node, and Common.js interfaces. 
A small subset of these is summarized in Fig.~\ref{tab:vitro}:
  attempts to access the file-system, exfiltrate data through the network, overwrite the module cache, access environment variables, and read the process arguments.

\sys infers two permissions: \ttt{\{e: \{eval: RX, exports: W\}\}}.
Having \sys enforce these permissions allows only primitive operations within \ttt{e}, such as arithmetic on primitive values and basic string manipulation.
It disallows all other operations with side-effects outside \ttt{e}, mitigating these attacks.

\subsection{Efficiency and Scalability (Q4)}
\label{q4}

\heading{Static Analysis}
To evaluate the efficiency and scalability of the static analysis, we run it on \topThousandpkgsToConsider{} npm packages gathered from the most depended-upon packages and the 50 benchmark libraries described above. In total, the analyzed packages comprise \topThousandSLOC{} LoC.
The static analysis successfully analyses all packages without errors, except for malformed files on which the Google Closure Compiler without our analysis also fails, \eg files containing syntax errors. 

The analysis takes only \topThousandAnalysisTime{} in total, \ie an average of \topThousandAnalysisTimePerLib{} per npm package.
To put these results into context, we also measure the performance of a popular linter---a lightweight static analysis pass flagging common human errors.
We use \ttt{eslint} (v5.0.1)~\cite{eslint} and each library's default linting configuration, falling back to Google's style rules when no such configuration exists.
\ttt{eslint} ranges between 0.94--6.017$s$ across the set of our benchmark libraries, averaging 1.34$s$ per library.
These results show that the static inference scales well to real-world modules and that its efficiency is comparable to other tools that developers use regularly.

 \begin{table}[t]
 \center
 \footnotesize
 \caption{
   Number of inferred \R, \W, \X, and \I permissions. 1\textsuperscript{st}Q, 2\textsuperscript{nd}Q and 3\textsuperscript{3r}Q represent the first quartile, the median, and the third quartile.
 }
 \vspace{-1mm}
 \begin{tabular}{llllll}
   \toprule
    Type       & Min & 1\textsuperscript{st}Q & 2\textsuperscript{nd}Q & 3\textsuperscript{3r}Q & Max                \\
   \midrule
    \R & \topThousandReadsMin &\topThousandReadsFirstQ & \topThousandReadsMedian &  \topThousandReadsThirdQ & \topThousandReadsMax\\
    \W & \topThousandWritesMin &\topThousandWritesFirstQ & \topThousandWritesMedian &  \topThousandWritesThirdQ & \topThousandWritesMax\\
    \X & \topThousandExecutesMin &\topThousandExecutesFirstQ & \topThousandExecutesMedian &  \topThousandExecutesThirdQ & \topThousandExecutesMax\\
    \I & \topThousandImportsMin &\topThousandImportsFirstQ & \topThousandImportsMedian &  \topThousandImportsThirdQ & \topThousandImportsMax\\
    \midrule
    Total & \topThousandTotalPermsMin &\topThousandTotalPermsFirstQ & \topThousandTotalPermsMedian &  \topThousandTotalPermsThirdQ & \topThousandTotalPermsMax \\
     \bottomrule
 \end{tabular}
 \label{tab:permsSAatScale}
 \end{table}

On average, the analysis infers one permission per \topThousandPermPerLOC{} lines of library code
In Table~\ref{tab:permsSAatScale}, we show statistics about the different types of permissions inferred by \sys. Most of them are \ttt{R} and \ttt{X} permissions (\topThousandPercReads{} and \topThousandPercExecutes{}, respectively), showing that the client packages rarely write to references outside of their boundaries, \eg to global variables or the API of the packages they use. While this is not entirely surprising, it shows that the probability of certain types of errors, \eg concurrent writes to the same global variable name, as addressed by ConflictJS~\cite{conflictjs}, is low for npm packages. The fact that a typical package in our experiments only imports six other entities is somehow consistent with the previous findings of Zimmermann et al.~\cite{npmstudy:19} that show that the direct number of dependencies is relatively low for npm packages, but their transitive number is high due to the fragmentation of the ecosystem.

\heading{Dynamic Enforcement}
Fig.~\ref{fig:dperf} shows the performance of \sys's runtime enforcement component for the 50 library benchmarks, as compared with with the performance of the unmodified runtime (V8 baseline).

\sys's full \rwx enforcement adds between 0.13--4.14$ms$ slowdown to executions that range between 324$ms$ and 2.77$s$.
Slowdowns average about 3.3$ms$ per library causing 1.93\% slowdown.
Based on these results, we do not anticipate a need for users to trade in runtime security to gain performance.

To better understand the sources of these overheads, we perform a series of micro-benchmarks with tight loops calling several ES-internal libraries without any I/O.
The primary source of overhead is the use of JavaScript's \ttt{with} construct~\sx{core}, 
The \ttt{with} construct dominates overheads, as it (1) interposes on too many accesses, only a fraction of which are relevant, and (2) remains significantly unoptimized, since its use is strongly discouraged by the JavaScript standards~\cite{with}.

\section{Related Work}
\label{related}
\new's techniques touch upon a great deal of previous work in several distinct domains.

\heading{Privilege Reduction}
A number of works have addressed privilege reduction~\cite{multics, rushby1981design, accetta1986mach, provos2003preventing, privman:03, privtrans:04, bittau2008wedge, wu2012codejail, salus:13, nodesentry:14, gudka2015clean, melara2019enclavedom}, often offering significant automation.
This automation often comes at the cost of \emph{lightweight annotations} on program objects---\eg configurations in Privman~\cite{privman:03}, \texttt{priv} directives in Privtrans~\cite{privtrans:04}, tags in Wedge~\cite{bittau2008wedge}, and compartmentalization hypotheses in SOAAP~\cite{gudka2015clean}.
TRON~\cite{tron} introduced a permission model similar to \sys, but at the level of processes rather than libraries.

Wedge and SOAAP stand out as offering some automation via dynamic and static analysis, respectively.
However, Wedge still requires altering programs manually to use its API, and SOAAP mostly checks rather than suggests policies.
In comparison to these works, \sys (1) leverages existing boundaries, and (2) offers significantly more automation.

To ameliorate manual annotations on individual objects, more recent library-level compartmentalization~\cite{breakapp:ndss:2018, sandcrust, pyronia:19} exploits runtime information about module boundaries to guide compartment boundaries.
These systems automate the creation and management of compartments, but do not automate the specification of policies through some form of inference.
\sys 
(1) focuses on benign-but-buggy libraries, rather than actively malicious ones, and 
(2) offers a simplified \rwx permission model rather than more expressive (often Turing-complete) policies---both in exchange for significant automation in terms of the permissions.

Pyxis~\cite{cheung2012automatic} and PM~\cite{liu2019program} reduce the problem of boundary inference to an integer programming problem by defining several performance and security metrics.
These systems are complementary to \sys, as they focus on separating the application code into a sensitive and insensitive compartment to minimize these metrics, while \sys tries to automatically infer and restrict the permissions between different libraries.

The static permission inference of \sys relates to work on statically inferring permissions that an application requires in the Java permission model~\cite{Koved2002}.
Jamrozik et al.~\cite{Jamrozik2016} describe a dynamic analysis to infer pairs of Android permissions and UI events that trigger the need for a permission.
We rely on static inference instead, to avoid the problem of automatically exercising the analyzed code.
An important difference to both above approaches is that \sys focuses on \ttt{RWX} permissions for specific access paths, instead of the more coarse-grained permissions supported by Java and Android.

\heading{Language-based Isolation}
\sys draws inspiration from language-based isolation techniques~\cite{sfi, aiken2006deconstructing}:
  rather than depending on the operating system for protection, \sys enforces access control from with the language runtime---even for libraries that access the file-system and the network.
Software fault isolation~\cite{sfi} modifies object code of modules written in \emph{unsafe} languages to prevent them from writing or jumping to addresses outside their domains.
Singularity's software-isolated processes~\cite{aiken2006deconstructing} ensure isolation through verification.
Leveraging memory safety, \new can be applied in environments with \emph{runtime code evaluation}, for which verification and static transformation might not be an option.

\heading{JavaScript Isolation}
There is significant prior work on JavaScript protection~\cite{mickens2014pivot, miller2008safe, terrace2012javascript, agten2012jsand, COWL}, motivated by the pervasive use of multi-party mashups on the client.
\sys is unusual in its model, inference, and enforcement:
  (1) it only allows simple \rwx permissions rather than Turing-complete functions;
  (2) its inference is static rather than dynamic; %
  (3) it leverages the library-import mechanism to apply both source- and context-transformations, thus wrapping a library's full observable behavior rather than just its interface.

NodeSentry~\cite{nodesentry:19} proposes powerful policies mediating between libraries.
Different from \sys, these policies are Turing-complete, are applied manually, and focus on inter-library APIs rather than a library's full observable behavior.

\heading{Capabilities}
Capability systems~\cite{levy84capability, shapiro1999eros} and object-capability systems~\cite{capolicies:13, miller2008safe} place access restrictions by restricting the ability to \emph{name} a resource, essentially treating the object reference graph as an access graph.
To make capabilities benefits more widespread, efforts such as Joe-E for Java~\cite{joe-e:10} and Caja for JavaScript~\cite{miller2008safe} restrict popular languages to object-capability-safe subsets.
Similar to capability systems, \sys interferes with the program's ability to name a resource---but rather than disabling naming, it augments it with a permission check.
\sys does not focus on a language subset, and its static-analysis offers significant automation.

\heading{Formalization}
Prior work has developed formal frameworks for stating and proving strong isolation properties in the context of new languages or subsets of existing languages~\cite{rscc, formalization, dimoulas2014declarative, drops:18, melicher2020controlling}.
\sys takes a different approach, working with an existing language and offering a quantification of privilege reduction rather than an all-or-nothing property.
\sys's \rwx model can be seen as overlaying a type system over the base language semantics, albeit one that is simple and geared towards existing codebases.

\heading{Software De-bloating}
Functionality elimination~\cite{rinard2011manipulating} and, more recently, software de-bloating~\cite{heo2018effective, koo2019configuration, azad2019less} lower potential vulnerabilities by eliminating unused code in a program.
These techniques focus on making subvertible, dormant code inactive---thus are similar in terms of goal to \sys, but approach the problem differently.
Rather than eliminating code, \sys makes the code inaccessible at runtime.

\heading{Ecosystem-focused}
Non-academic responses to the challenges of third-party libraries~\cite{nodesecurity, retireJS, snyk} focus on tools that check the program's dependency chain for known vulnerabilities.
Dependency locking~\cite{shrinkwrap} does not rule out security problems;
  on the contrary, users forego valuable bug and vulnerability fixes, while experiencing a more convoluted dependency management.
Deno, a new JavaScript/TypeScript runtime, offers a coarse-grained allow-deny permission model focusing on the file-system and the network~\cite{deno}.
While this addition confirms that practitioners are still lacking a solution, it lacks \sys's advanced automation and fine granularity.

\section{Conclusion}
\label{conclusion}

Dynamic library compromise due to bugs or other problems---even when libraries themselves are not malicious---poses a serious security threat in modern software development.
\sys addresses it by augmenting the module system with a fine-grained read-write-execute (\rwx) permission model that allows specifying privilege at the boundaries of libraries.
\sys performs an analysis to infer these permissions to not overburden developers and then enforces them using a lightweight load-time code transformation.
Finally this paper proposes privilege reduction, a metric that can be used to quantitatively evaluate \sys's impact on prevented permissions. 

This paper shows that \sys manages to prevent real attacks without breaking library functionality while (i) requiring minimal user effort, and (ii) imposing minimal overhead on execution and compilation time. 
\sys is therefore meant to be a lightweight addition to any contemporary developer's toolkit, complementing---rather than replacing---other defense mechanisms.
\sys is available for system-wide installation via \ttt{npm}~[blinded], and its source code is available on GitHub:
\begin{lstlisting}[numbers=none]
        github.com/blindorg/mir
\end{lstlisting}

\bibliographystyle{plain}
\bibliography{../bib}

\appendix

\section{Zooming into Incompatibilities}
\label{two}

We now zoom into the sources of incompatibility between the inferred and required permissions, assuming no import-time inference---\ie between static analysis and dynamic semantics.
\begin{figure}[t]
  \centering
  \includegraphics[width=\columnwidth]{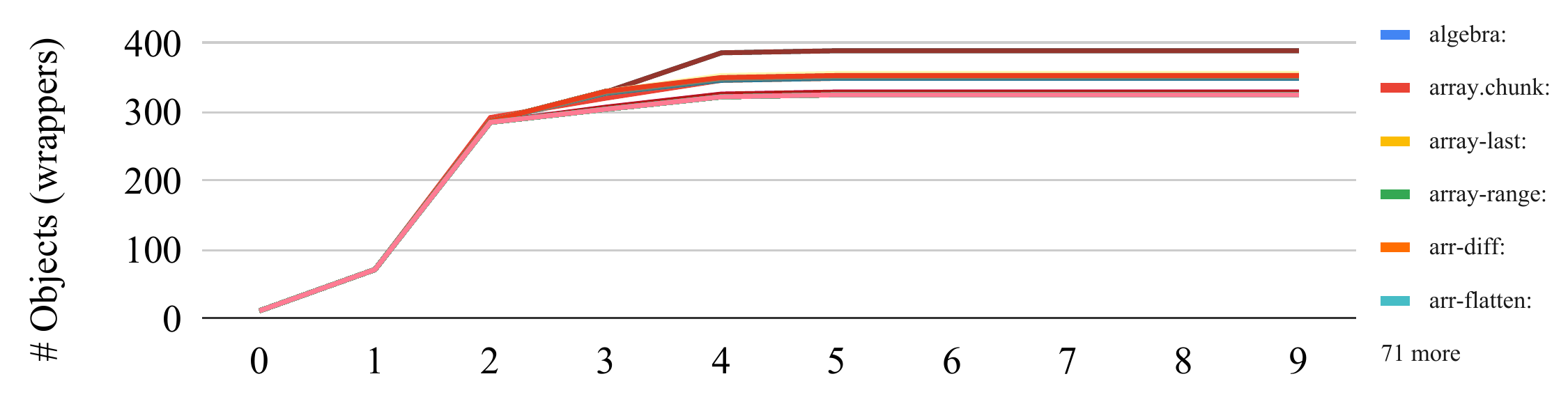}
  \includegraphics[width=\columnwidth]{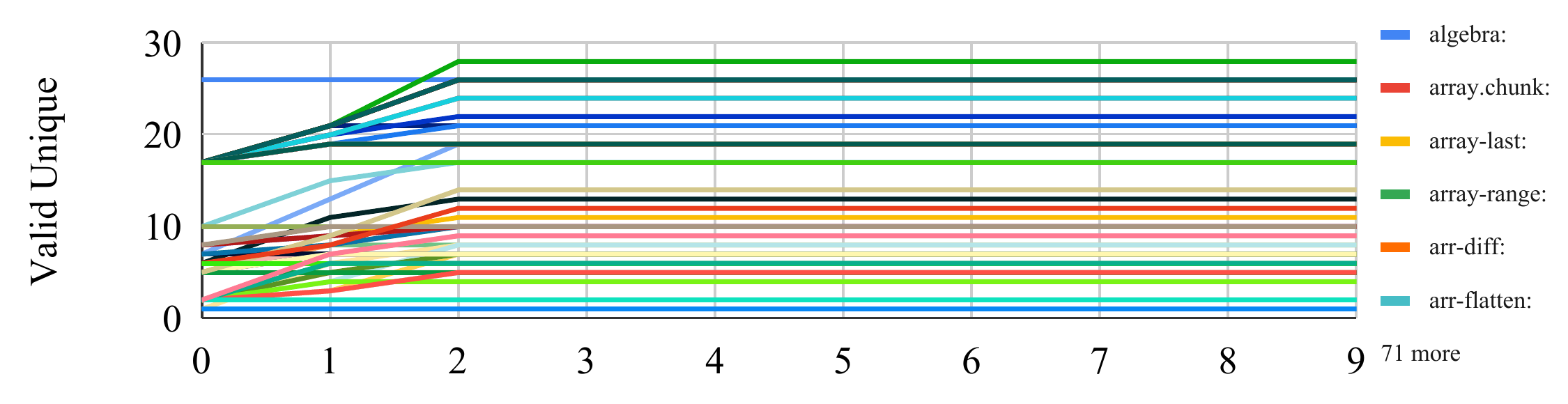}
  \includegraphics[width=\columnwidth]{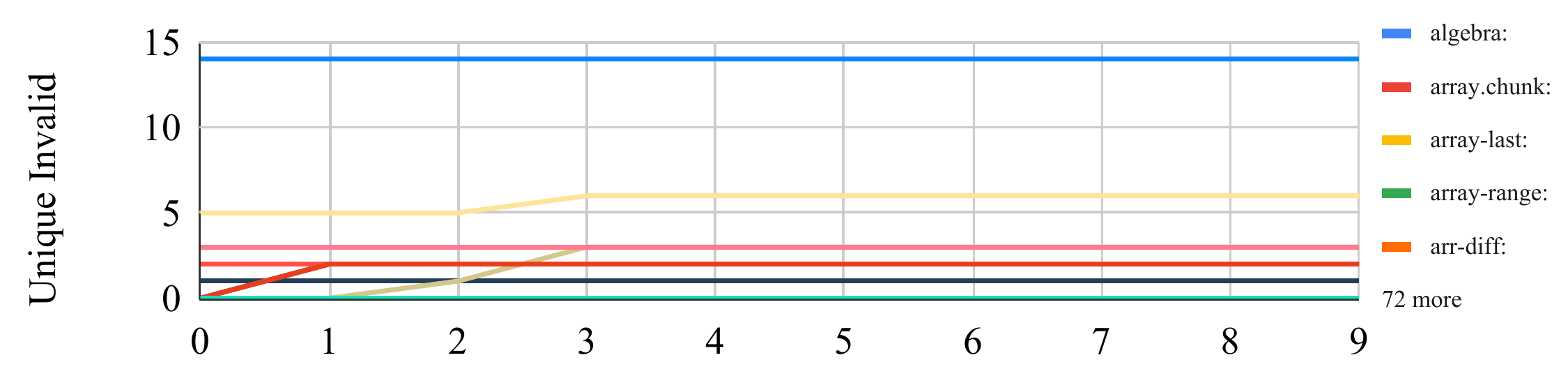}
  \includegraphics[width=\columnwidth]{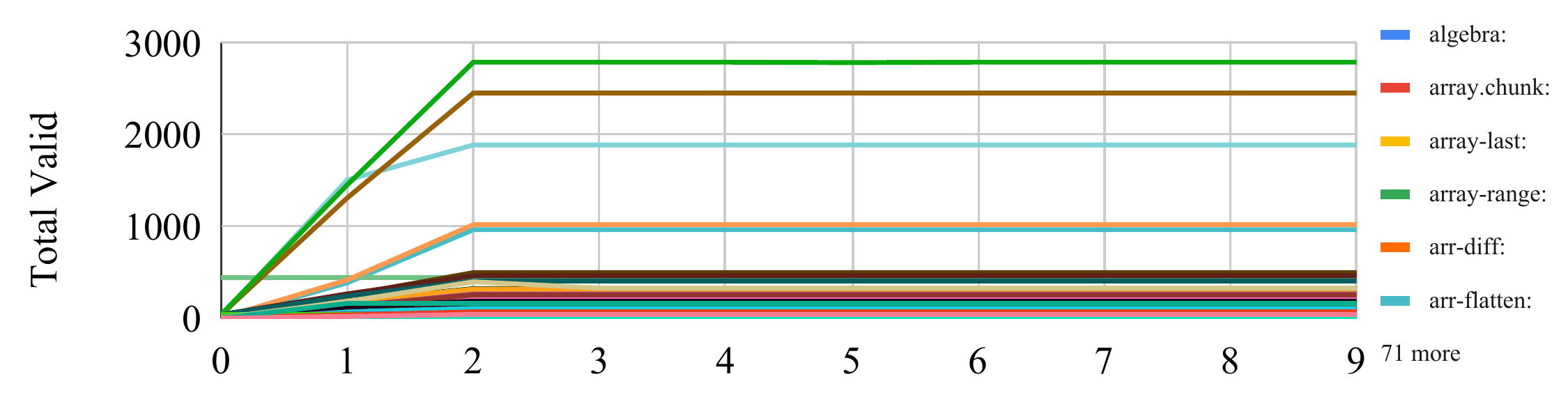}
  \includegraphics[width=\columnwidth]{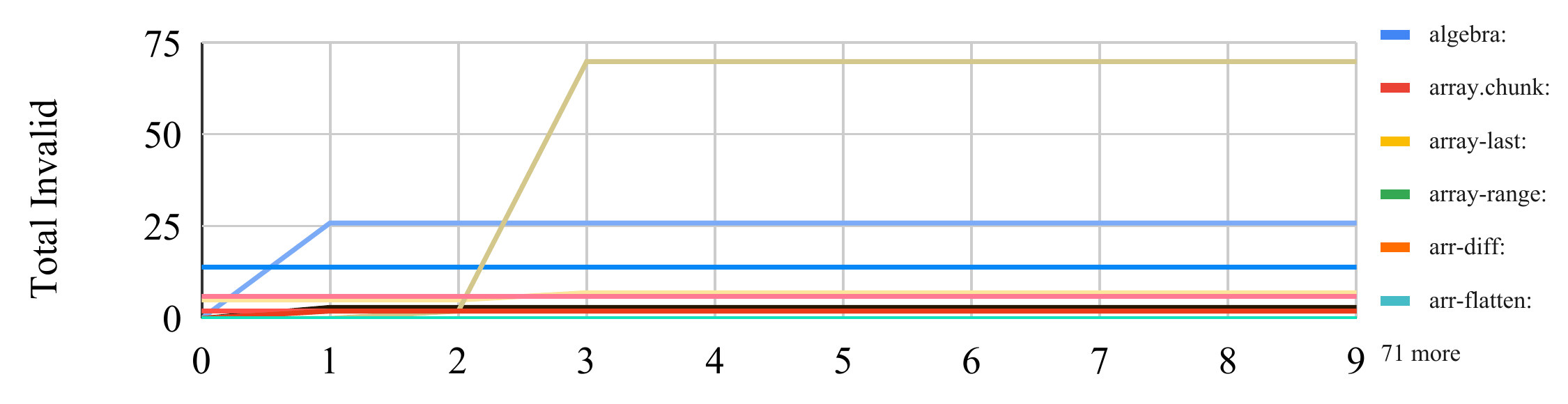}
  \caption{
    \textbf{Depth vs. compatibility.}
    From the top:
    (a) number of \sys wrappers,
    (b) unique valid accesses,
    (c) unique invalid accesses,
    (d) total valid accesses, and
    (e) total invalid accesses---all as a function of  depth~\cf{q1}.
  }
  \label{fig:objects}
\end{figure}

\begin{figure*}
  \centering
  \includegraphics[width=0.05\textwidth]{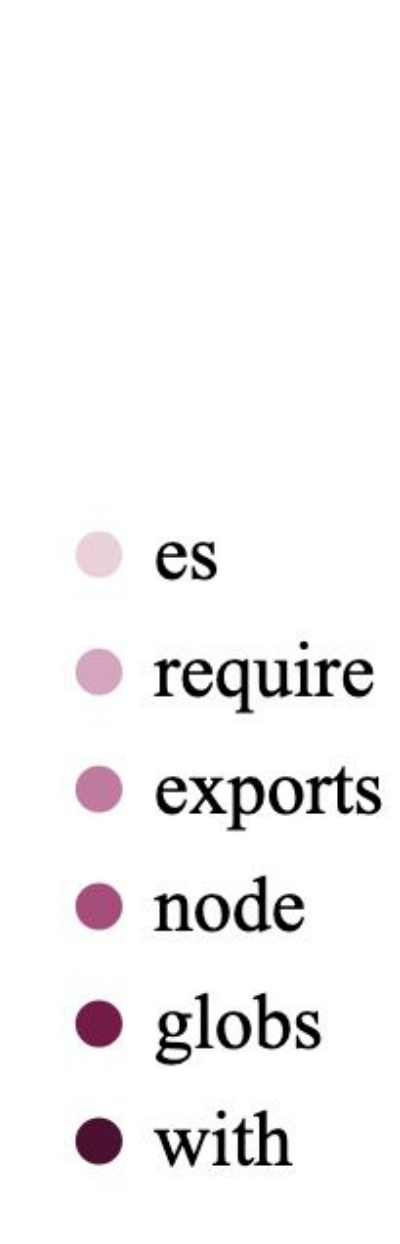}
  \includegraphics[width=0.18\textwidth]{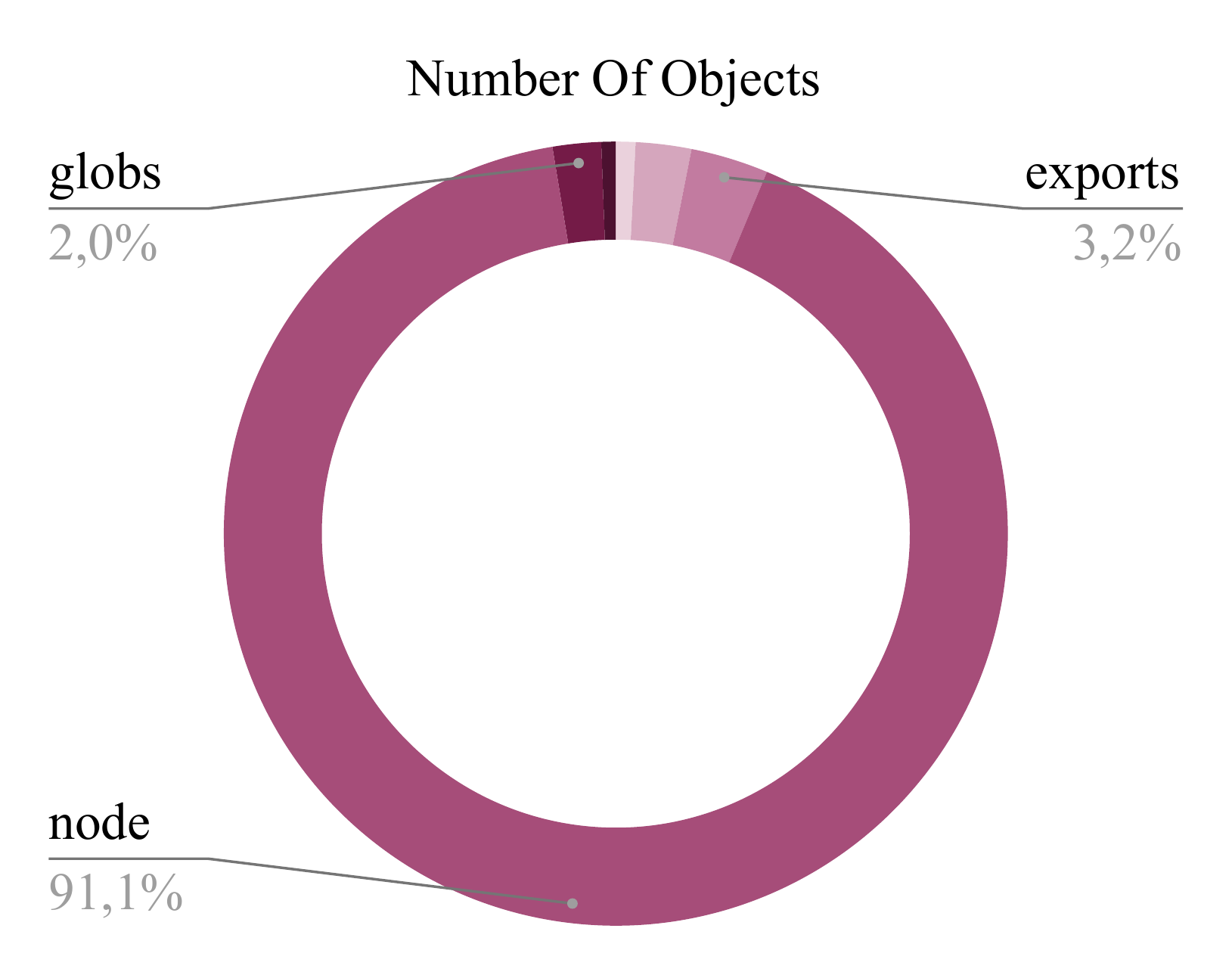}
  \includegraphics[width=0.18\textwidth]{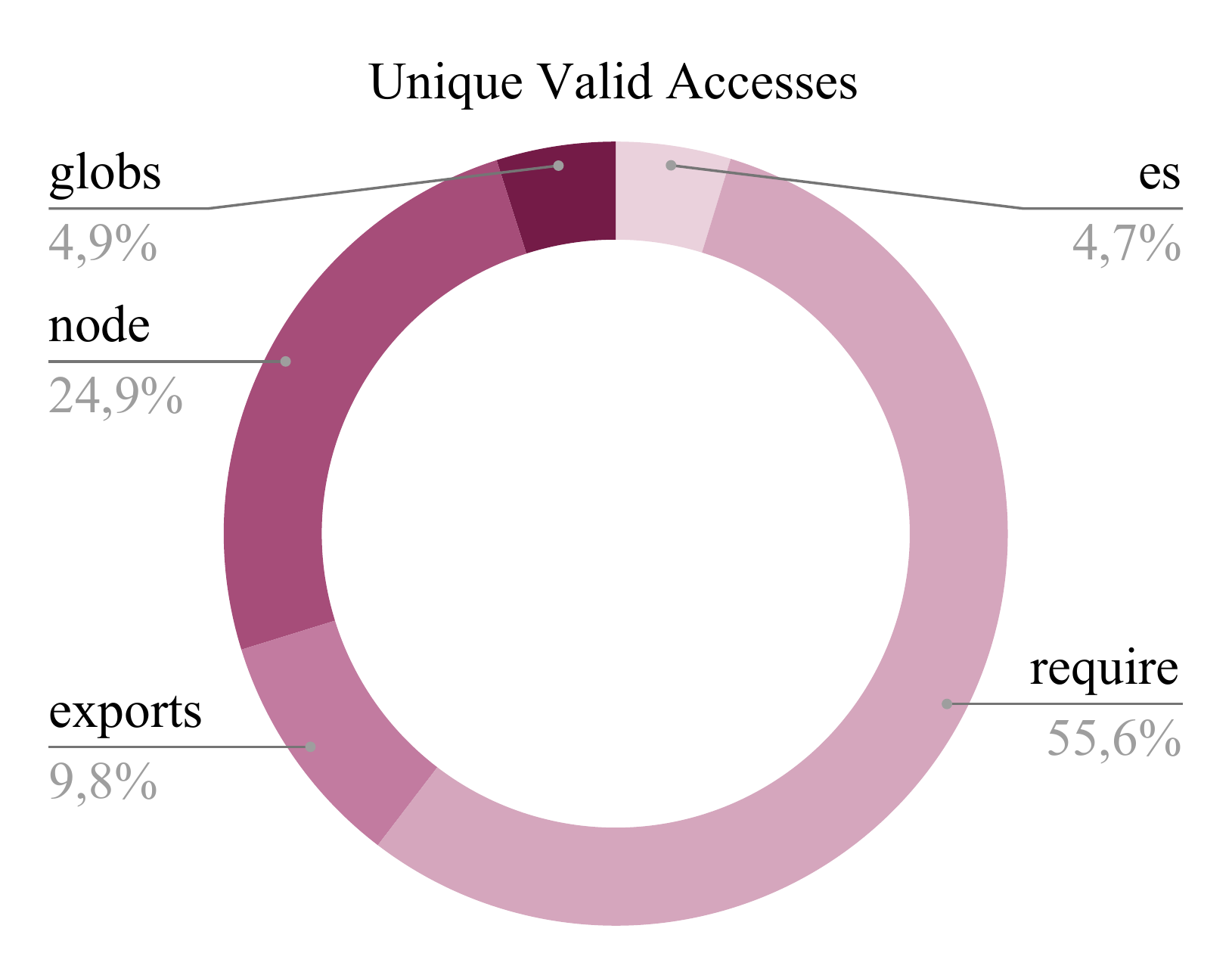}
  \includegraphics[width=0.18\textwidth]{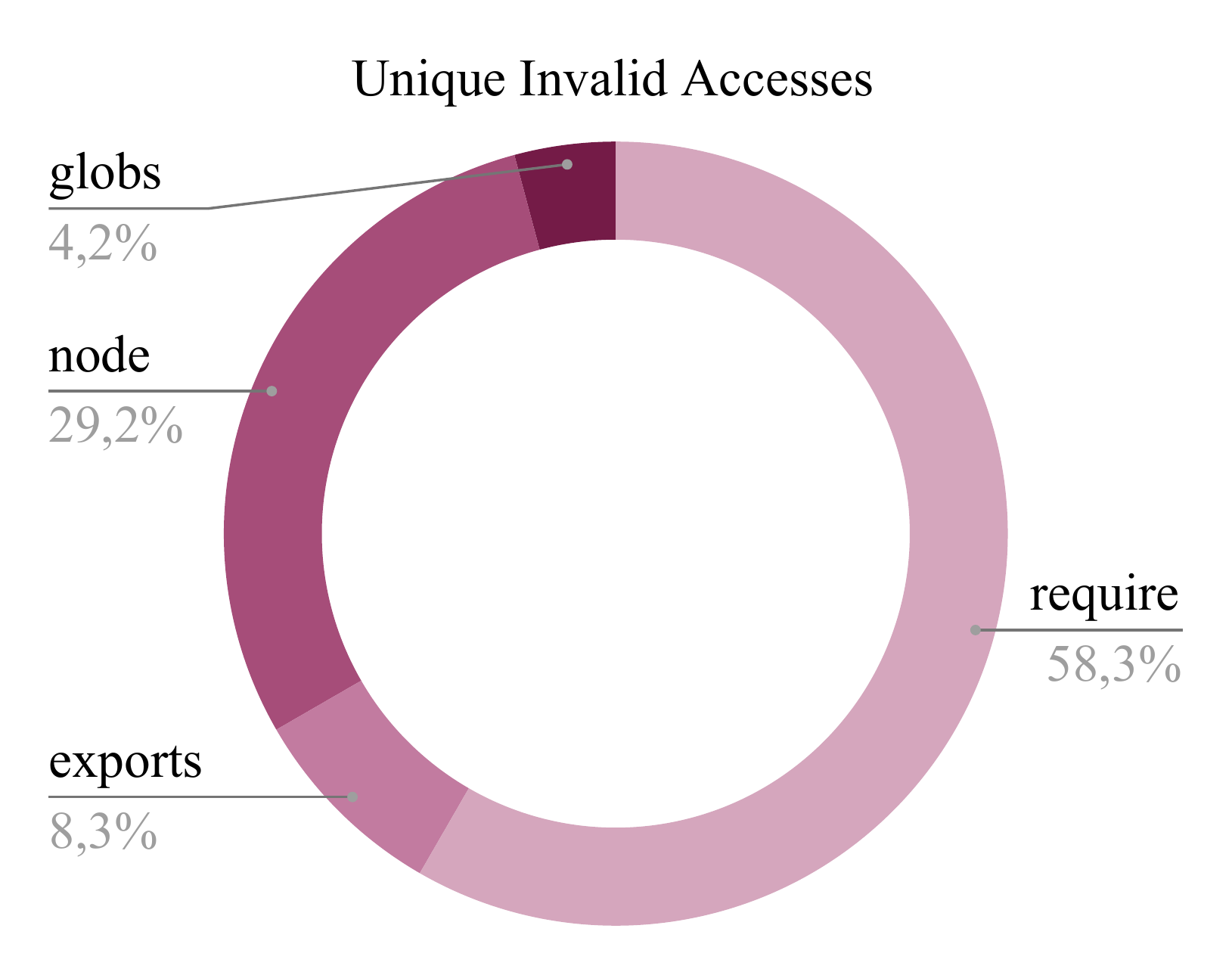}
  \includegraphics[width=0.18\textwidth]{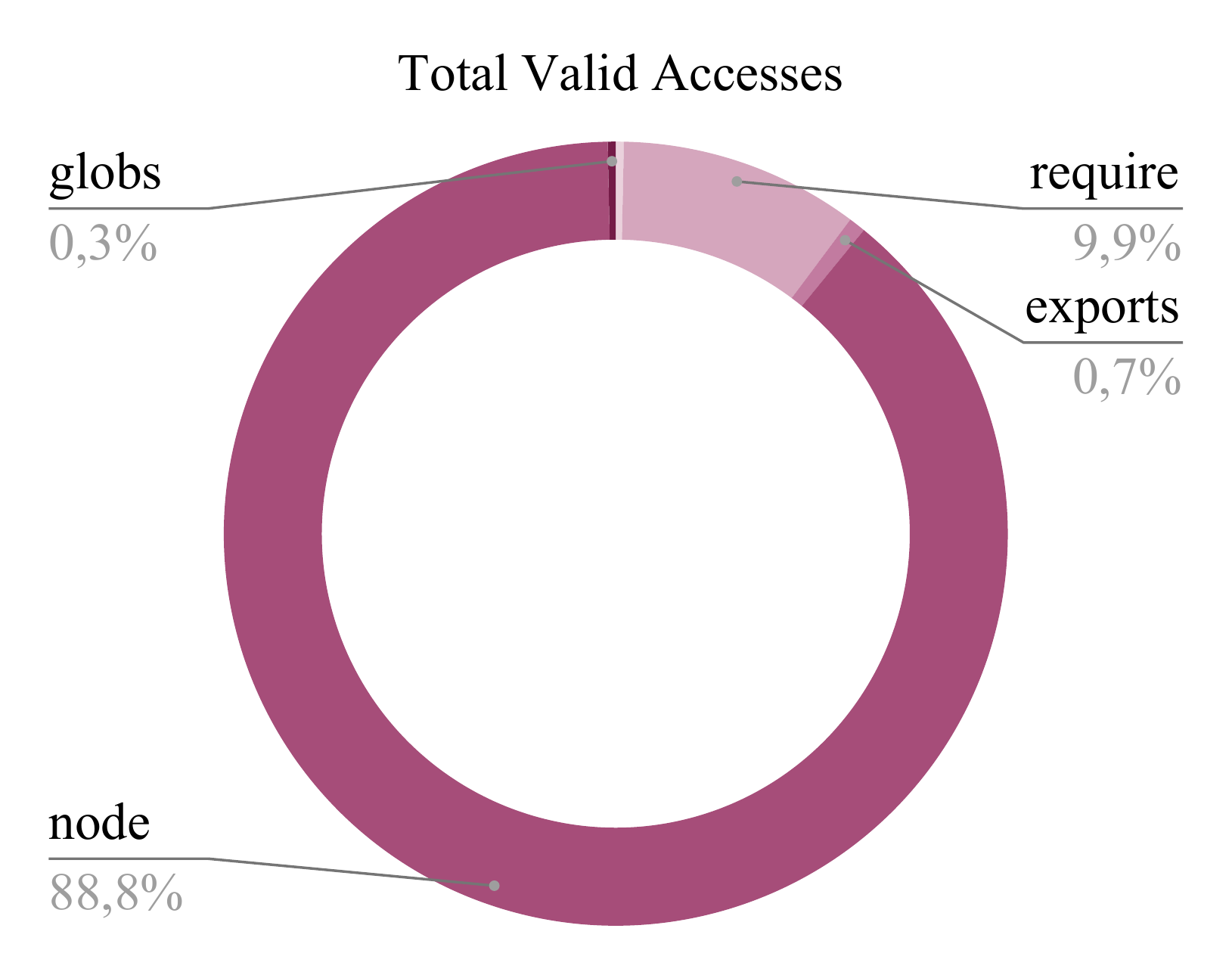}
  \includegraphics[width=0.18\textwidth]{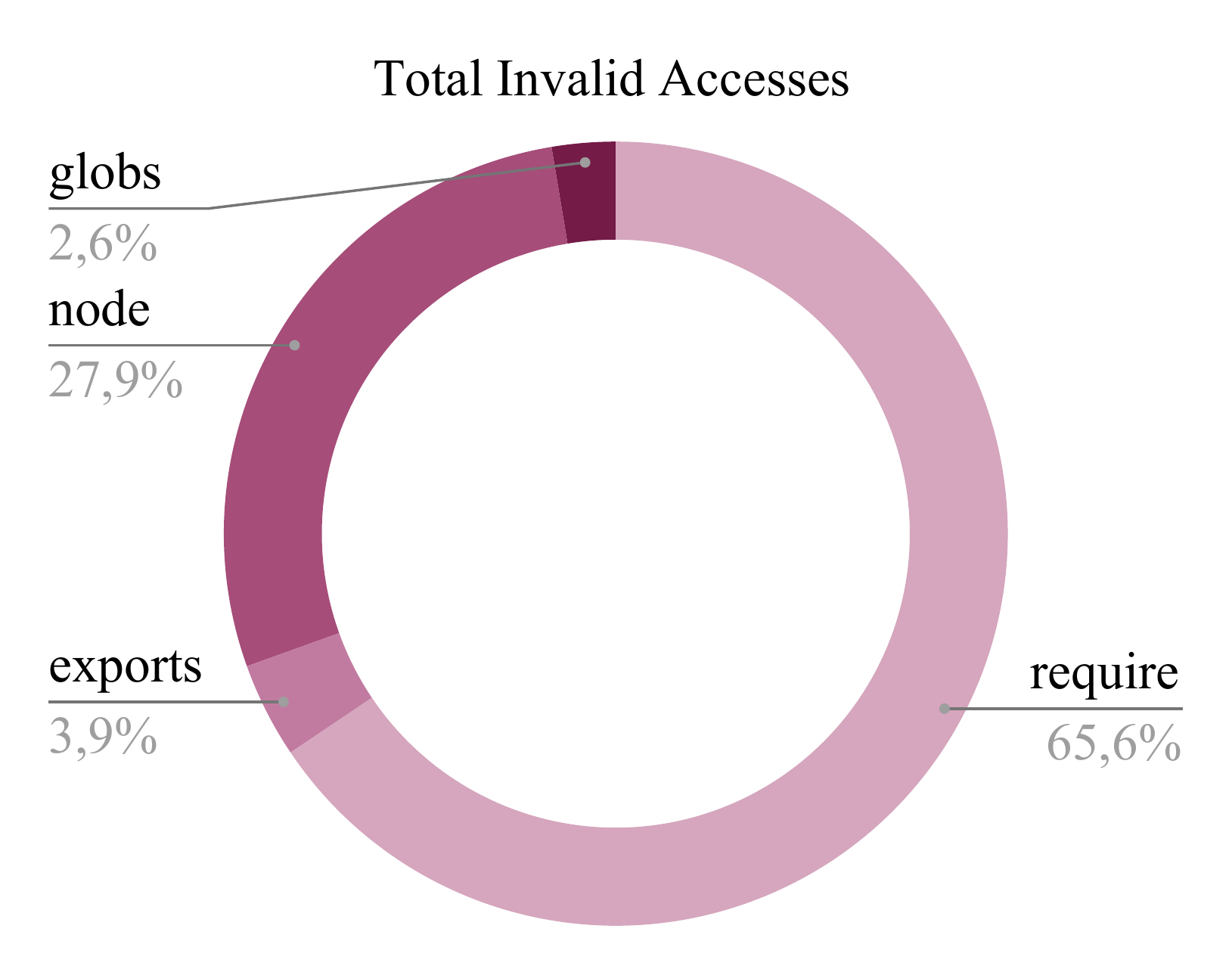}
  \caption{
    \textbf{Context vs. compatibility.}
    From the left:
    (a) number of \sys wrappers,
    (b) unique valid accesses,
    (c) unique invalid accesses,
    (d) total valid accesses, and
    (e) total invalid accesses---all as a function of context~\cf{q1}
  }
  \vspace{-5mm}
  \label{fig:context}
\end{figure*}

\heading{Context}
Context refers to the broad source of names that are available in the current scope~\sx{contracts}.
This includes names
  defined by the EcmaScript standard (\ctx{es}), 
  through an explicit import (\ctx{exports}),
  by the Node.js runtime (\ctx{node}), or
  via global variables (\ctx{globs}).
A few names seem globally available but are in fact module-locals (\ctx{require}).
User-defined global variables are not prefixed with \ttt{global} thus requiring special interposition. %

Fig.~\ref{fig:context} shows several statistics across all 50 libraries as a function of context.
In terms of the number of objects wrapped, the vast majority comes from \ctx{node} which accounts for 91.1\% of all wrappers.
In terms of unique number of accesses, for both valid and invalid ones the majority comes from \ctx{require}.
Taking the number of accesses into account, the majority of accesses comes from \ctx{es} and \ctx{node} names, whereas the majority of invalid ones comes from \ctx{exports}.

These results highlight that \sys gathers most of its correct results around \ctx{node} and \ctx{es} names.
In retrospect, this makes sense:
  there is no dynamic resolution required to see, for example, that \ctx{Math} refers to the standard math library---and this is the most common to refer to it.
The majority of invalid accesses come from module exports.
This is expected, as a few meta-programming libraries use introspection and reflection capabilities to rewrite entire interfaces and re-export them dynamically.

\heading{Depth}
Depth refers to how far \sys traverses references starting from the names of objects in  the contexts.
For example, the access \ctx{global}.\fld{obj}.\fld{x} is two levels deep and \ctx{fs}.\fld{readFile} is one level deep.

Fig.~\ref{fig:objects} shows the same statistics as before but relative to depth.
There are a few highlights worth noting.
First, while the number of objects which \sys's runtime component keeps track of starts growing exponentially, it quickly reaches an average upper bound of about 400.
Second, accesses grow quickly for the first couple of levels---objects for some of these libraries are hit several thousand times---but then remain stable.
This is usually because many interfaces follow a mostly-flat format where all methods are defined in the top level.
Lastly, the majority of invalid accesses remain constant across several levels of depth.
This is mostly because these libraries exhibited dynamic behavior at exactly one point---\eg the \ttt{export} object or a dynamically-computed property (see Tab.~\ref{tab:reasons}).

\end{document}